\documentclass[a4paper,12pt]{article}  
\pdfoutput=1
\usepackage{amsmath,amsfonts,amssymb,epsfig,comment,xspace,listings,hyperref,cite}
\usepackage{graphicx,caption,subcaption}

\numberwithin{equation}{section}
\usepackage{slashed}

\usepackage{youngtab}
\usepackage{listings}
\usepackage{xcolor}
\usepackage{booktabs}

\usepackage{tabularx, multirow}

\makeatletter

\def\pd[#1]{\frac{\partial}{\partial #1}}
\def\pdd[#1,#2]{\frac{\partial #1}{\partial #2}}

\def\beq{\begin{equation}}
\def\eeq{\end{equation}}

\def\twomat[#1,#2][#3,#4]{\left( \begin{array}{cc} #1 & #2 \\ #3 & #4 \end{array} \right)}
\def\twoa[#1,#2][#3,#4]{\left( \begin{array}{cc} #1 & #2 \\ #3 & #4 \end{array} \right)}

\def\thv[#1,#2,#3]{\left( \begin{array}{c} #1 \\ #2 \\ #3 \end{array} \right)}
\def\twv[#1,#2]{\left( \begin{array}{c} #1 \\ #2 \end{array} \right)}

\def\SARAH{{\tt SARAH}\xspace}
\def\SPheno{{\tt SPheno}\xspace}
\def\MicrOmegas{{\tt MicrOmegas}\xspace}
\def\HiggsSignals{{\tt HiggsSignals}\xspace}
\def\HiggsBounds{{\tt HiggsBounds}\xspace}
\def\Vevacious{{\tt Vevacious}\xspace}
\def\Vevaciouspp{{\tt Vevacious++}\xspace}
\def\Flexible{{\tt FlexibleSUSY}\xspace}

\definecolor{maroon}{cmyk}{0, 0.87, 0.68, 0.32}
\definecolor{halfgray}{gray}{0.55}
\definecolor{slha_frame}{RGB}{207, 207, 207}
\definecolor{slha_bg}{RGB}{247, 247, 247}
\definecolor{slha_red}{RGB}{186, 33, 33}
\definecolor{slha_green}{RGB}{0, 128, 0}
\definecolor{slha_cyan}{RGB}{64, 128, 128}
\definecolor{slha_purple}{RGB}{170, 34, 255}

\definecolor{mathematica_frame}{RGB}{207, 207, 207}
\definecolor{mathematica_bg}{RGB}{247, 247, 247}
\definecolor{mathematica_red}{RGB}{186, 33, 33}
\definecolor{mathematica_green}{RGB}{0, 128, 0}
\definecolor{mathematica_cyan}{RGB}{64, 128, 128}
\definecolor{mathematica_purple}{RGB}{170, 34, 255}

\lstnewenvironment{MIN}[1][]{%
  \renewcommand{\thelstnumber}{In[\arabic{lstnumber}]}
  \lstset{language=MathIn,numbers=left,basicstyle=\ttfamily,#1}%
}{%
}

\lstnewenvironment{MOUT}[1][]{%
  \renewcommand{\thelstnumber}{Out[\arabic{lstnumber}]}
  \lstset{language=MathOut,numbers=left,basicstyle=\ttfamily,#1}%
}{%
}

\usepackage{listings}
\lstdefinelanguage{SLHA}{
    morekeywords={block,Block,BLOCK,decay,Decay,DECAY},%
    sensitive=true,%
    morecomment=[l]\#,%
    morestring=[b]',%
    morestring=[b]",%
    morestring=[s]{'''}{'''},
    morestring=[s]{"""}{"""},
    morestring=[s]{r'}{'},
    morestring=[s]{r"}{"},%
    morestring=[s]{r'''}{'''},%
    morestring=[s]{r"""}{"""},%
    morestring=[s]{u'}{'},
    morestring=[s]{u"}{"},%
    morestring=[s]{u'''}{'''},%
    morestring=[s]{u"""}{"""},%
    identifierstyle=\color{black}\ttfamily,
    commentstyle=\color{slha_cyan}\ttfamily,
    stringstyle=\color{slha_red}\ttfamily,
    keepspaces=true,
    showspaces=false,
    showstringspaces=false,
    rulecolor=\color{slha_frame},
    frame=single,
    frameround={t}{t}{t}{t},
    framexleftmargin=6mm,
    numbers=left,
    numberstyle=\tiny\color{halfgray},
    backgroundcolor=\color{slha_bg},
    basicstyle=\footnotesize,
    keywordstyle=\color{slha_green}\ttfamily,
    aboveskip=1.2em,
    belowskip=1.2em,
}

\lstdefinelanguage{MathIn}{
    morekeywords={Simplify,Eigenvalues},%
    emph={Start,InitUnitarity,GetScatteringDiagrams,BuildScatteringMatrix,MakeSPheno},%
    emphstyle={\color{mathematica_purple}},
    sensitive=true,%
    morecomment=[l]\%,%
    morestring=[b]',%
    morestring=[b]",%
    morestring=[s]{'''}{'''},
    morestring=[s]{"""}{"""},
    morestring=[s]{r'}{'},
    morestring=[s]{r"}{"},%
    morestring=[s]{r'''}{'''},%
    morestring=[s]{r"""}{"""},%
    morestring=[s]{u'}{'},
    morestring=[s]{u"}{"},%
    morestring=[s]{u'''}{'''},%
    morestring=[s]{u"""}{"""},%
    identifierstyle=\color{black}\ttfamily,
    commentstyle=\color{mathematica_cyan}\ttfamily,
    stringstyle=\color{mathematica_red}\ttfamily,
    keepspaces=true,
    showspaces=false,
    showstringspaces=false,
    rulecolor=\color{mathematica_frame},
    frame=single,
    frameround={t}{t}{t}{t},
    framexleftmargin=10mm,
    numbers=left,
    numberstyle=\tiny\color{halfgray},
    backgroundcolor=\color{mathematica_bg},
    basicstyle=\footnotesize,
    keywordstyle=\color{mathematica_green}\ttfamily,
    aboveskip=1.2em,
    belowskip=1.2em,
}

\lstdefinelanguage{MathOut}{
    morekeywords={Simplify,Eigenvalues},%
    sensitive=true,%
    morecomment=[l]\%,%
    morestring=[b]',%
    morestring=[b]",%
    morestring=[s]{'''}{'''},
    morestring=[s]{"""}{"""},
    morestring=[s]{r'}{'},
    morestring=[s]{r"}{"},%
    morestring=[s]{r'''}{'''},%
    morestring=[s]{r"""}{"""},%
    morestring=[s]{u'}{'},
    morestring=[s]{u"}{"},%
    morestring=[s]{u'''}{'''},%
    morestring=[s]{u"""}{"""},%
    identifierstyle=\color{black}\ttfamily,
    commentstyle=\color{mathematica_cyan}\ttfamily,
    stringstyle=\color{mathematica_red}\ttfamily,
    keepspaces=true,
    showspaces=false,
    showstringspaces=false,
    rulecolor=\color{mathematica_frame},
    frame=single,
    frameround={t}{t}{t}{t},
    framexleftmargin=10mm,
    numbers=left,
    numberstyle=\tiny\color{halfgray},
    backgroundcolor=\color{mathematica_bg},
    basicstyle=\footnotesize,
    keywordstyle=\color{mathematica_green}\ttfamily,
    aboveskip=1.2em,
    belowskip=1.2em,
}

\let\origthelstnumber\thelstnumber
\makeatletter
\newcommand*\Suppressnumber{%
  \lst@AddToHook{OnNewLine}{%
    \let\thelstnumber\relax%
     \advance\c@lstnumber-\@ne\relax%
    }%
}

\newcommand*\Reactivatenumber{%
  \lst@AddToHook{OnNewLine}{%
   \let\thelstnumber\origthelstnumber%
   \advance\c@lstnumber\@ne\relax}%
}

\numberwithin{equation}{section}

\title{Active learning BSM parameter spaces}

\date{}

\begin{document}

\begin{flushright}
\end{flushright}
\begin{center}

\vspace{1cm}
{\bf \LARGE Active learning BSM parameter spaces}

\vspace{1cm}

\large{Mark D. Goodsell\footnote{goodsell@lpthe.jussieu.fr} and
Ari Joury\footnote{ari@joury.eu}
 \\[5mm]}

{ \sl Laboratoire de Physique Th\'eorique et Hautes Energies (LPTHE),\\ UMR 7589,
Sorbonne Universit\'e et CNRS, 4 place Jussieu, 75252 Paris Cedex 05, France.}

\end{center}
\vspace{0.7cm}

\abstract{Active learning (AL) has interesting features for parameter scans of new models. %
We show on a variety of models that AL scans bring large efficiency gains to the traditionally tedious work of finding boundaries for BSM models. %
In the MSSM, this approach produces more accurate bounds. %
In light of our prior publication
, we further refine the exploration of the parameter space of the SMSQQ model, and update the maximum mass of a dark matter singlet to 48.4 TeV. %
Finally we show that this technique is especially useful in more complex models like the MDGSSM.
}

\newpage
\setcounter{footnote}{0}


\section{Introduction}
\label{sec:intro}

The Standard Model (SM) is a remarkably economic theory, making many predictions based on only a few parameters. In principle all of the parameters of the model can be fixed by observations, and then many more observables can be predicted. Once we add new particles and interactions to the SM -- the bread and butter of many theoretical particle physicists -- we necessarily add many new parameters. It is then in general a very difficult task to fix as many of these by existing observations and then make predictions for new observations. In many cases these observations have not yet been made; for example, we usually cannot fix the mass of a particle we have not yet observed. In others, the relationship with the observable is complicated; for example the Higgs mass in supersymmetric models, which is highly sensitive to loop corrections (see e.g. \cite{Slavich:2020zjv}). Therefore the first task with a new model is generally to explore the parameter space for plausible allowed regions according to some basic list of observations, before a more detailed examination can be made.

There now exists a whole chain of tools for computing the properties of general new theories. Most relevant for this work are \SARAH\cite{Staub:2008uz,Staub:2013tta,Goodsell:2015ira,Goodsell:2015yca,Braathen:2017izn,Goodsell:2018tti,Goodsell:2020rfu}, \SPheno\cite{Porod:2003um,Porod:2011nf} and \MicrOmegas\cite{Belanger:2018ccd,Belanger:2020gnr}, but there exist many others, such as \HiggsSignals\cite{Bechtle:2013xfa,Bechtle:2020uwn}, \HiggsBounds\cite{Bechtle:2008jh,Bechtle:2020pkv}, \Vevacious/\Vevaciouspp\cite{Camargo-Molina:2013qva} and \Flexible\cite{Athron:2014wta,Athron:2017fvs}.
Thanks to SUSY Les Houches Accords \cite{Skands:2003cj,Allanach:2008qq}, the output of these codes is standardised and can be passed from one to the other in the form of text files. However, the actual task of doing this job is generally left to the user, and of course there is then the task of choosing the strategy of exploring the parameter space of models.

Almost universally the community has adopted the practice of computing a likelihood function based on the desired set of observables and using techniques developed in other fields to sample the parameter space with a point density proportional to this likelihood. In its simplest form, this involves assigning a gaussian or log likelihood to each observable, with a given mean and variance that may be related to experimental measurements and uncertainties (e.g. when selecting for observables such as $\Delta \rho$), or just reflect the uncertainty of the tools (e.g. taking an uncertainty of the Higgs mass of the order a few GeV). These likelihoods are generally combined assuming no correlations. The search strategy is then often a Markov Chain Monte Carlo (MCMC) based on the Metropolis-Hastings alogrithm; or other groups use more efficient versions based e.g on {\tt MultiNest} \cite{Feroz:2008xx}. 

{\tt GAMBIT} \cite{GAMBIT:2017yxo,GAMBITModelsWorkgroup:2017ilg}, especially with the advent of the GAMBIT Universal Machine \cite{Bloor:2021gtp}, attempts to solve this problem for the user by creating backends for common tools and taking over the task of interfacing between them. Likelihoods are computed by its included tools and combined across all observables; and the user then has the option of certain included search strategies \cite{Martinez:2017lzg}. This is an admirable and powerful tool and best for large-scale scans on a computing cluster. 

One problem with the use of likelihood functions is that for many use cases in High Energy Physics (HEP) they vary sharply around a small or narrow region in high-dimensional parameter spaces. While in principle they are continuous functions and therefore contain useful information when exploring regions away from the experimentally allowed region, allowing a smart algorithm to guide itself towards more likely points, in practice, and especially when combining many unrelated observables, they are effectively step or delta functions.  This was noted in \cite{Ren:2017ymm}, where a different approach, named ``Machine Learning Scan,'' was proposed in and later implemented in the tool {\tt xBit} \cite{Staub:2019xhl}. This involves training a neural network directly on observables rather than on a likelihood function and this was then used to select points that would be in the ``good'' region. The user can then reasonably rapidly find many points in the allowed region, again with sample density proportional to the likelihood in that region, reproducing the results of MCMC-based strategies after sampling fewer points.  

In this work we are interested in a different goal, namely for users who merely want to find, as a first step, the \emph{boundaries} of the allowed region. Indeed, for many phenomenologists exploring a model for the first time, the most likely point or region is profoundly uninteresting; for many models they are more likely if we set new physics effects to large masses or very weak couplings. As a more pertinent example, in our previous work \cite{Goodsell:2020rfu} we were interested in finding the \emph{heaviest possible} dark matter mass for a given model; this is a long way away from the most likely region, and we were interested there in exploring the decision boundary itself. There we adapted a simple MCMC algorithm by biasing the likelihood function to favour heavier masses. Here we shall revisit that computation and show more clearly the effect of such a bias on the sampling distribution. However, the main result of this work is a new algorithm using \emph{active learning} to train a neural network discriminator and \emph{choose points near the decision boundary} so as to best explore the limits of the allowed parameter space of a model.

Active Learning (AL) (for a useful review see \cite{SettlesReview}) is the general name for machine learning where the algorithm chooses its own inputs. This means finding a measure of the uncertainty of the algorithm about its prediction for any given inputs, and then choosing points where the uncertainty is greatest. Since a neural network used as a predictor does not have a natural definition for this uncertainty, this is generally applied to approaches such as random forest classifiers, where the decision is made based on the results of a \emph{set of models}, and then the uncertainty can be related to the differing predictions within the set. This has recently been applied in the HEP context\cite{Caron:2019xkx,Rocamonde:2022gyw}. Here we note that a neural network \emph{classifier} does have a natural notion of uncertainty, and we propose an algorithm to use this to efficiently select points in high-dimensional parameter spaces. Since neural networks are much more sophisticated than random forests with unlimited potential for generalisation, for more complicated parameter spaces this should lead to more efficient sampling and discovery, and -- despite the extra overhead in training neural networks -- save time and give a more accurate description of the allowed range of a model. Indeed, once the model has been trained through active learning, the discriminator can then be used to describe the parameter space.

This paper is organised as follows. In section~\ref{sec:algorithm} we describe our algorithm and setup. We then apply AL scans to several models of increasing complexity: %
first, simple toy models (see section~\ref{sec:toy}), then the CMSSM (section~\ref{sec:mssm}), the SMSQQ~\cite{Goodsell:2020rfu} (section~\ref{sec:smsqq}), and finally the MDGSSM (section~\ref{sec:mdgssm}). %
We also compare the performance to standard MCMC scans, and to vanilla neural networks and random forest classifiers (RFCs). %
We conclude with a brief discussion of the possible use cases of AL scans, and perspectives for future work.
The code for this work is implemented in a general framework and so will be released publically along with an upcoming publication.

\section{Active learning with a neural network}
\label{sec:algorithm}

In this section we describe our AL algorithm to train a neural network. We will assume the reader is familiar with neural networks and deep learning; they are becoming ubiquitous in science,  see e.g.~\cite{Abdughani:2019wuv,Boehnlein:2021eym,Gili:2022oul} for recent HEP theory applications. 

Our approach is based on data that can be simply categorised as ``good'' or ``bad.'' This decision can be made on a given point by an ``oracle'' which in a simple model could just be a formula, whereas in HEP applications it will be based on ranges of observables computed using given tools. In traditional likelihood-based approaches, being somewhat outside of the ``excluded'' range for one observable could be compensated by being very likely in other obserables, and indeed this has a reasonable explanation that if we sample hundreds of observables randomly then we should expect a few to be outside their ``allowed'' range. In our approach it is for the user to decide what ``allowed'' or interesting range they want to investigate, and this gives added flexibility without introducing biases.

Our first step is to create an initial dataset, typically consisting of random points, and query the oracle to get the result. 
We then feed these points and the oracle results to a neural network and let it train on those. We describe the parameters of the networks used in the subsequent sections, but they all consist of an input layer feeding to a series of $2$ to $5$ hidden layers of large (order $100$) size connected by {\tt ReLU} activation functions, and a final ouput layer with one neuron whose output is mapped to a sigmoid function. The implementation is done in {\tt pytorch} and we use the default weight initialisation. 

During training the neural network tries to minimize the loss function; we use the Binary Cross Entropy:  
\begin{align}
\text{loss} = -\frac{1}{N}\sum_{i=1}^{N} y_i \log (\hat{y}_i) + (1-y_i) \log (1-\hat{y}_i) ,
\label{eq:bsmart:scans:nn:formulas:node2}
\end{align}
where $y_i$ is the outcome of the data, and $\hat{y}_i$ the fitted outcome by the neural network. %
In the case of what is studied in this work, $\hat{y}_i$ is some number between 0 and 1, so we use a sigmoid function for the output of the final layer); %
$y_i$ is either 0 or 1 depending whether a point is bad or good.
This loss function is minimized by influencing $\hat{y}_i$ by changing the weights of the networks. %


After initial training, at each further step in the scanning/training cycle, our aim will be to choose $K$ points to pass to the oracle, from $L$ proposed to the discriminator. The results of the oracle's evaluation of the $K$ points are then used to further train the discriminator. However, we have several choices about both how to propose the $L$ points, and then how to select the batch of $K$. An initial strategy would be to choose the $L$ points completely randomly (similar to the MLS approach\cite{Ren:2017ymm}); this would help to randomly discover interesting regions, but is very inefficient in high-dimensional parameter spaces. Indeed, if we have $n$ parameters with an interesting region a hypercube of side $0.1$ of the parameter range, then a random scan would require $\mathcal{O} (10^n)$ points to find it; if $L=100000$ and $n\ge 6$, then subsequent passes would not be likely to pick a point near the region. As an alternative we can choose points ``near'' good points that we have found, somewhat like the jumps during an MCMC; however, in that case we do not want to spend time in uninteresting regions away from the boundary near many good points, if the ``good'' region is large. Therefore we adopt a hybrid approach of choosing $10\%$ of the $L$ points purely randomly, and the remaining $90\%$ from the vicinity of good points. 

Furthermore, if the discriminator is not providing useful information then it is hardly worthwhile choosing points near its decision boundary. Hence the batch of $K$ points contains a further proportion $p$ of the $K$ points selected randomly/within jumps of good points. 
We choose $p$ based on the training score of the classifier: if $q$ is the proportion of points incorrectly classified after the last training, then
$$ p = 2 \times \mathrm{min}(q,0.5).$$
In other words, if the discriminator is no better than a coin toss then we need to propose entirely random points. 

In order to select the $pK$ points to pass to the oracle we also need a prescription for scoring the batch. First we assign an uncertainty score $s_i$ to each point:
\begin{align}
s_i = \hat{y}_i (1-\hat{y}_i).
\label{eq:code:al:kfroml:kfroml:score}
\end{align}
This score $s_i$ peaks at $\hat{y}_i=0.5$ and reaches 0.25 there, and falls off to 0 for $\hat{y}_i =0$ and $\hat{y}_i=1$. It is essentially equivalent to the Binary Cross Entropy above for one variable, but simpler to calculate. An initial choice would be to select the $pK$ points which score highest.
However, this would potentially lead to clustered points and insufficient diversity. This is also a classic issue in AL, and is solved by introducing a diversity/distance measure \cite{DiversityAL} to quantify the distance between points in feature space. In random forest appraoches, naively the entropy of the predictions on the sample can be maximised; or the Kullback-Leibler (KL) divergence can be used. In our case this is not possible, but we can instead use a score based on the physical distance in parameter space! We experimented with using a positive score for larger distances, but found that this simply drove the sampling to the boundaries, so instead we introduce a distance measure based on an electrostatic repulsion between points. If $\underline{x}_i, \underline{x}_j$ are the vectors representing the input parameters for two points, and $d^2 = |\underline{x}_i- \underline{x}_j|^2$ then the ``repulsion'' is given by
\begin{align}
r(\underline{x}_i , \underline{x}_j) = \left\{ \begin{array}{cl}- \frac{a}{a + d^2}, & d< 0.01 \\ 0 & d\ge 0.01 \end{array} \right.
\end{align}
Here $a$ is some constant that we take to be $0.0001$. Of course, this requires that we have rescaled all of our input parameters to have range $[0,1]$. We then start with the point that has the highest score $s_i$ and remove it from the pool $\overline{P}$ and put it in the selected set $\overline{pK}$. Then we iteratively add points until we have selected $pK$ points as follows:
\begin{enumerate}
\item Compute $r_j\equiv \sum_i r(\underline{x}_i , \underline{x}_j)$ for $\{\underline{x}_i\} \in \overline{pK}, \{\underline{x}_j\} \in \overline{P}$ for each point $ \underline{x}_j$
\item Compute the maximum total repulsion $r_{\rm \max} = \mathrm{max}(\{r_j\})$ and the standard deviation $\sigma$  of the uncertainty scores $\{s_i \}$.
\item Assign to each point a score:
  $$ S_i = (1-\alpha) s_i +\alpha\ r_i\ \sigma \times \frac{1}{4 r_\text{max}}$$
\item Add the point with the highest score $S_i$ to the set $\overline{pK}$ and remove it from the pool $\overline{P}$.
\end{enumerate}
Note that at each step it is not necessary to recompute the whole sum $r_j$, since by storing the old scores we just need to add the scores for the total repulsion from the last point we added to our selected set. 
The parameter $\alpha$ is a diversity weighting that can be adjusted depending on the scan: if we think that the sample contains only one small intersting region we may wish to set $\alpha$ small. By using the standard deviation and $r_{\rm max}$ we make sure that the relative weight of the diversity measure and the neural network uncertainty have a weight that depends only on $\alpha$.

Once we have our set of $K$ points, we pass these to the oracle, and then train the discriminator on the outputs over a given number of epochs. We also have a choice as to whether to train the discriminator on the \emph{whole} dataset or just on new points. The cost of training on the full dataset is time, especially once a large number of points have been accrued; however, only training on new data, especially when those points are chosen to be near the boundary, can be deleterious. Hence we train on the full dataset every fixed number of iterations of this procedure.

As a final remark, it is also necessary to \emph{balance} the dataset for training the discriminator: in cases where we have many bad points and few good ones (especially initially) it is advantageous for the discriminator to just classify everything as bad. Hence in our training set we make copies of the underrepresented point set so that the discriminator is fed an equal number of good and bad points at each training epoch.

\section{Active learning toy models}
\label{sec:toy}

\begin{figure}
\centering
\begin{subfigure}{.30\textwidth}
  \centering
  \includegraphics[width=\linewidth]{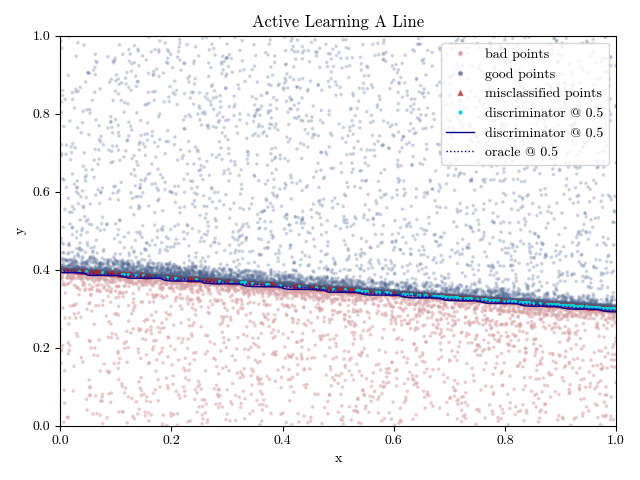}
\end{subfigure}%
\begin{subfigure}{.30\textwidth}
  \centering
  \includegraphics[width=\linewidth]{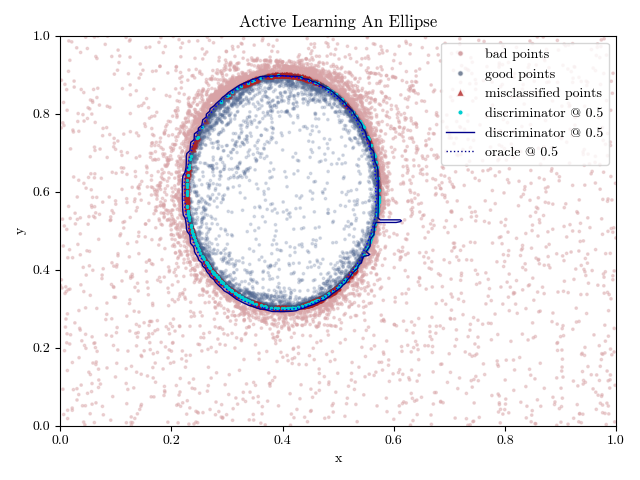}
\end{subfigure}%
\begin{subfigure}{.30\textwidth}
  \centering
  \includegraphics[width=\linewidth]{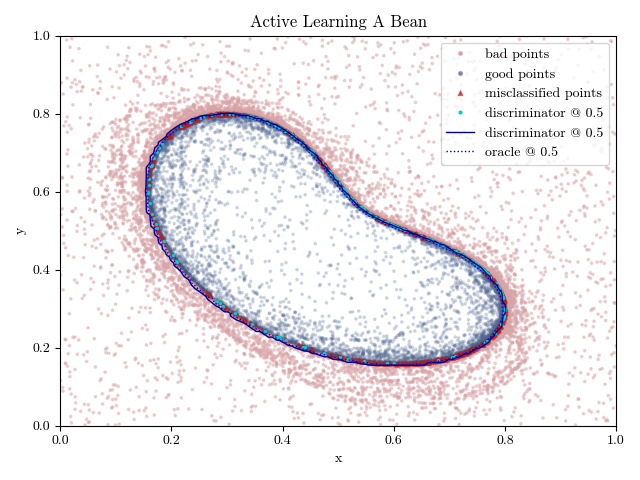}
\end{subfigure}
\begin{subfigure}{.30\textwidth}
  \centering
  \includegraphics[width=\linewidth]{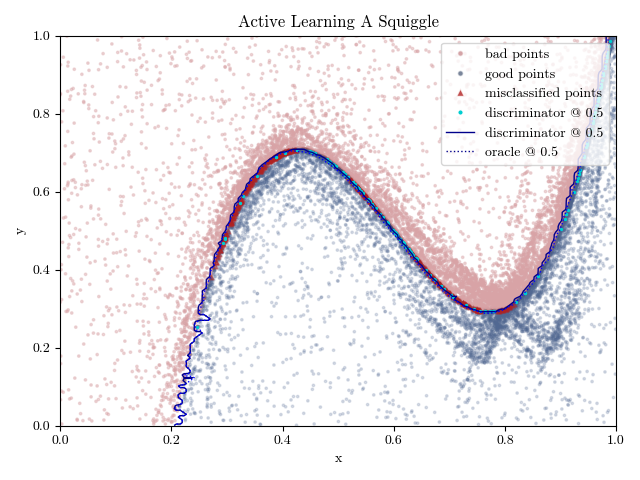}
\end{subfigure}%
\begin{subfigure}{.30\textwidth}
  \centering
  \includegraphics[width=\linewidth]{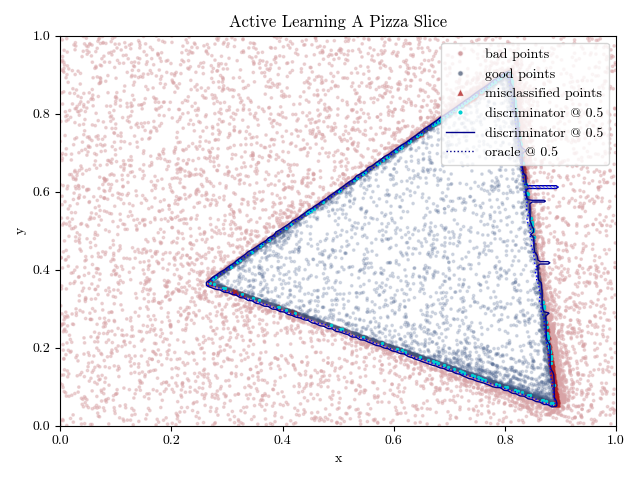}
\end{subfigure}%
\begin{subfigure}{.30\textwidth}
  \centering
  \includegraphics[width=\linewidth]{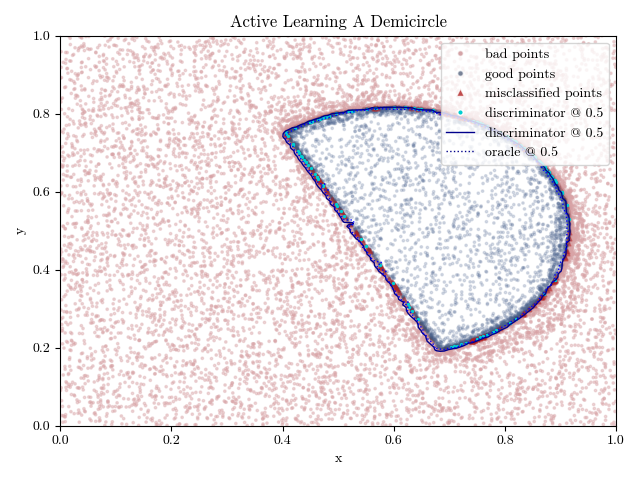}
\end{subfigure}
\caption{AL-generated points of various toy models with 20,000 points each. 
Note how well the discriminator (solid line) and the oracle (dotted line) match, even with pointy or irregularly shaped objects.
Also note that all points the discriminator is uncertain about or misclassified are on the oracle / discriminator lines.
}
\label{fig:toy:toy1}
\end{figure}

\begin{figure}
\centering
\begin{subfigure}{.45\textwidth}
  \centering
  \includegraphics[width=\linewidth]{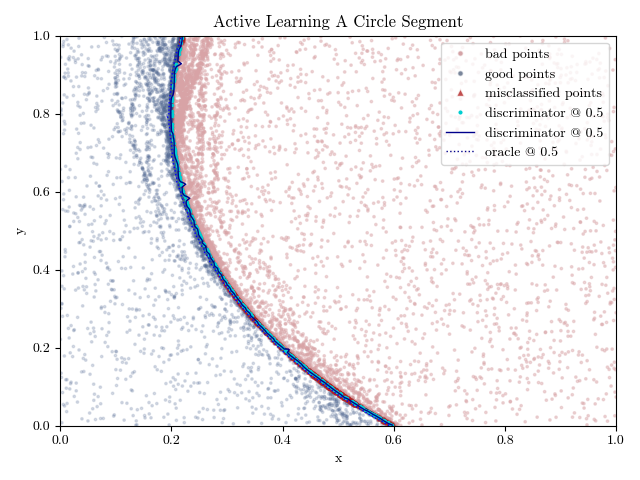}
\end{subfigure}%
\begin{subfigure}{.45\textwidth}
  \centering
  \includegraphics[width=\linewidth]{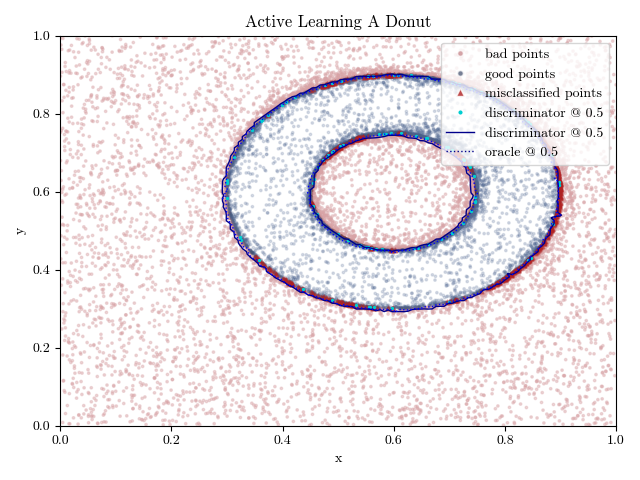}
\end{subfigure}
\begin{subfigure}{.45\textwidth}
  \centering
  \includegraphics[width=\linewidth]{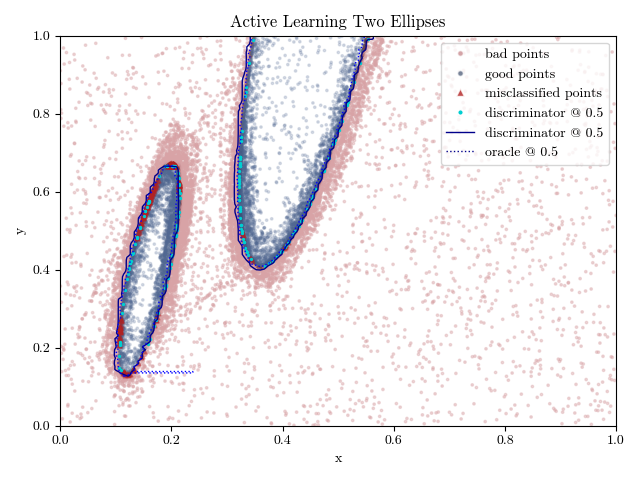}
\end{subfigure}%
\begin{subfigure}{.45\textwidth}
  \centering
  \includegraphics[width=\linewidth]{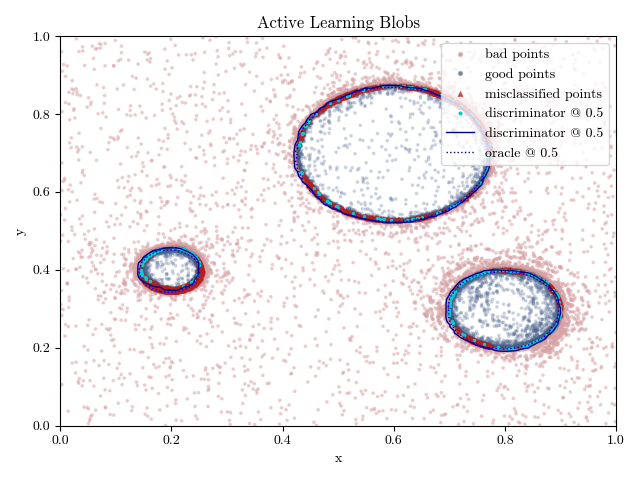}
\end{subfigure}
\caption{20,000 AL-generated points for more models.
Note how the oracle and the discriminator match snugly even around multiple interesting regions or regions with holes.
}
\label{fig:toy:toy2}
\end{figure}

We begin by illustrating the principle of active learning on a variety of toy models in two dimensions. %
The results of this are depicted in figures~\ref{fig:toy:toy1} and~\ref{fig:toy:toy2}. %
In these figures one can see a fair amount of randomly distributed points, which is intended in order to make sure that no potentially interesting region is missed. %
Crucially, though, we see how well the algorithm figuratively zooms into the interesting regions to determine where the border between good and bad points runs.

To obtain these results, we use a scan with settings as shown in table~\ref{tab:toy:settings}.
On all toy models, 20,000 points and 2,000 training steps suffice.
The settings work for a variety of shapes, including irregular ones like the bean and the squiggle (see figure~\ref{fig:toy:toy1}) %
and ones with holes or multiple pieces, like the beams and the blobs (see figure~\ref{fig:toy:toy2}).
Its ability to correctly identify multiple regions is particularly notable as this sets it somewhat apart from an MCMC scan.
The known danger in the latter is that it zooms into one good region and missing other good regions in the process.
This can be mitigated with additional measures.
But in active learning, this danger is circumvented automatically, so that parameter spaces with more than one good region are a lot easier to deal with.

\begin{table}[h!]
  \vspace{3mm}
  \footnotesize
  \centering
  \begin{tabularx}{0.88\textwidth}{|l|l|X|X|}
    \hline
    Setting & Value & Exceptions & Comments \\
    \hline
    \hline
    Number of points & 20,000 & -- & Total number of points per scan  \\
     & & & \\ 
    Initial random points & 1000 & Pizza, Donut: 5000; Demicircle: 8000 & Number of random points for initial training \\
     & & & \\
    K & 100 & -- & Number of points added per iteration \\
     & & & \\
    L & 5000 & -- & Number of random points from which the K most interesting ones are selected \\
     & & & \\
    Diversity Alpha & 0.005 & Pizza, Demicircle: 0.015 & Adjusts the distance of the K points \\
     & & & \\
    FullTrain & 8 & Donut, Demicircle: 4 & After this many iterations, retrain on whole dataset \\
     & & & \\
    Hidden Layers & 2 & -- & Number of hidden layers in the network \\
     & & & \\
    Hidden Size & 100 & -- & Number of nodes per hidden layer \\
     & & & \\
    Learning Rate & 0.001 & -- & How much the weights are changed in each training step \\
     & & & \\
    SGD momentum & 0.05 & -- & Stochastic gradient descent momentum, helps gradient vectors accelerate in the right direction \\
     & & & \\
    Training steps & 2000 & -- & Number of training steps \\
     & & & \\
    Weight decay & 0.001 & -- & Penalty to loss function, helps prevent overtraining \\
     & & & \\
    Epsilon & 0.3 & -- & The learning rate decreases by this factor if the training error goes up \\
    \hline
  \end{tabularx}
\caption{Network settings for the toy models.
Note that only 3 out of 10 models need additional finetuning, which speaks for the flexibility of our classifier.
}
\label{tab:toy:settings}
\end{table}

There are a few exceptions where the settings of our scan needed to be adjusted to account for special features of a model.
In the case of the donut, more initial random points were needed to identify the hole; %
without these random points the algorithm finds the outer border but misses the inner one. %
In the case of the pizza, one of the borders is not recognised without more initial random points. %
The straight line of the demicircle is not recognized with lower numbers of random points either. %

For the donut and the demicircle, the algorithm also needs to train twice as often on the full dataset in order to prevent it from weighting pieces of new data too much. %
The diversity alpha, which ensures that new points have a certain distance from one another, is also increased for the pizza and the demicircle. %
This helps identify the pointy corners of these shapes. %
For the donut, this is not necessary as has no pointy edges.

It is plausible, albeit far from proven, that similar settings adjustments may be necessary for models which exhibit holes or pointy edges in their parameter space. %
Nevertheless, the fact that we were able to correctly identify 7 out of 10 shapes with the exact same settings illustrates the versatility of active learning scans without the need for extensive finetuning.
\\ 

All figures show the results after 20,000 points for the sake of having a nice plot; %
however, the algorithm accurately identified all borders between good and bad points already after 10,000 points. %
It's quite plausible that an interpolation of a grid- or random scan with this amount of points could have led to a similar accuracy. %
Unless one can press this interpolation into an analytical form, however, the knowledge of the boundaries will be of limited use. %
In contrast, the active learning scan -- or any scan with a neural network, for this point -- returns a model in additon to a list of points. %
This allows us to do two additional things: %
First, we can query the model about another point, even if we already ran the scan and the point was not included in it. %
Second, we can refine this model by training it on more points if we so desire. %
Both these things would not be as straightforward with an interpolation. 

The active learning approach trumps other approaches employing neural networks in the sense that it finds points, unlike for example a vanilla neural network. %
In addition, it \textit{automatically} finds interesting points, i.e. points close to the boundaries. %
This puts it in a similar category as a GAN~\cite{Goodfellow:2014upx,Mohamed:2016xyz}, where two neural networks try to fool one another with increasingly interesting points. 
The advantage of our approach, however, is its relative simplicity.
With active learning, we only have one model's hyperparamters to tune -- finetuning the two competing networks of a GAN is notoriously difficult~\cite{Lopez:2022lkd,Mescheder:2018xyz} -- and we need to train less, meaning that we potentially use less compute.
And despite the simplicity, we find compelling accuracies even in higher-dimensional models, as can be seen in sections~\ref{sec:smsqq} and~\ref{sec:mdgssm}.


\section{Active learning in the CMSSM}
\label{sec:mssm}

Having tested our algorithm on toy oracles, we will now apply it to a simple physics example: the valid dark matter parameter space of the constrained MSSM. This is a scenario where the masses and gauge couplings unify at the GUT scale, and leads to only five free parameters called $m_0, m_{1/2}, A_0, \tan \beta$ and $\mathrm{sign}(\mu)$; it is therefore used as a standard theoretical example, especially for exploration of the valid dark matter parameter space, e.g.\cite{Ellis:2002wv,Ellis_2009,Ellis:2012aa}. To make the problem resemble the toy model we will only consider the $m_{0}/m_{1/2}$ plane, fixing $\tan \beta =10, A_0 =0$ and $\mathrm{sign}(\mu) = 1$. We shall scan over $m_0, m_{1/2} \in [100\, \mathrm{GeV},2000\, \mathrm{GeV} ]$ for both an example MCMC scan and our active learning algorithm (using the same codes of \SARAH and \MicrOmegas for both to give a fair comparison).

This is the same parameter plane considered in \cite{Staub:2019xhl} and in one example in \cite{Ellis:2012aa}. As in the former reference we only impose a constraint on the dark matter density and ignore all other constraints. The difference between those two references is in the lattitude allowed; the latter takes \mbox{$\Omega h^2 = 0.112 \pm 0.012$} while in the former $\Omega h^2 < 0.2$ is considered acceptable. We consider \mbox{$\Omega h^2 < 0.12$} to be a ``good'' point and larger densities to be ``bad.'' In the MCMC scan we use a log likelihood for $\Omega h^2$ with mean $0.112$ and variance $0.05$. This is peaked around the decision boundary and therefore should provide a simple alternative to our AL procedure to find points near it. In the AL scan we classify valid points as those with $\Omega h^2 < 0.12$. We use $L=10000,K=100$ and a variance of $200$ in the steps around good points. 

With those constraints, the parameter plane is not especially interesting: there is an acceptable region at very small $m_{1/2}$ which eventually leads to a region at $m_0 > 1250$ GeV for $m_{1/2}$ up to $400$ GeV where no points are generated by the spectrum generator because there is no electroweak symmetry breaking. On the left side of the plane at small values of $m_0$ and large $m_{1/2}$ there is a coannihilation region, but in reference  \cite{Ellis:2012aa} (and as we find with our constraint on $\Omega h^2$) this disappears into a region roughly from $(m_{1/2}, m_0) $ from $(100,540)\, \mathrm{GeV}$, up to $(350,1500)\, \mathrm{GeV}$ where the LSP is a charged slepton.

Hence in our training we ignore the constraint on the charged LSP and just take the dark matter density from \MicrOmegas -- which shows the density of the neutralino, overlaying the unphysical region on the plot aferwards. Since the points are not especially physical and the idea is to compare strategies, and with the results of the previous references this can therefore be regarded as a toy model. 



\begin{figure}
\centering
\begin{subfigure}{.45\textwidth}
  \centering
  \includegraphics[width=\linewidth]{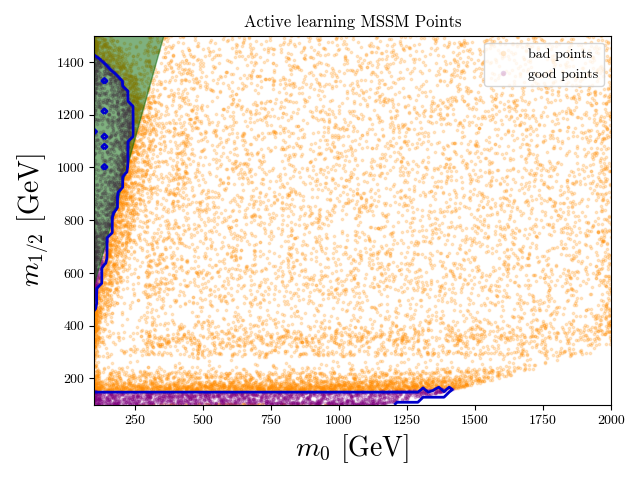}
\end{subfigure}
\begin{subfigure}{.45\textwidth}
  \centering
  \includegraphics[width=\linewidth]{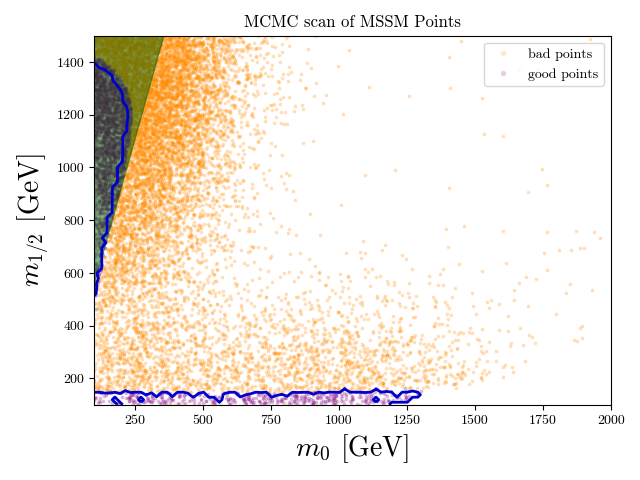}
\end{subfigure}
\caption{Point distribution in the $m_0-m_{12}$ plance of the MSSM with 20,000 points each. %
Top panel: AL scan; bottom panel: MCMC scan. %
The blue line indicates the location where the disciminator puts the line between good and bad points, as evaluated on a grid with 100 divisions in either direction. The dark green shaded area is the region with a charged LSP, which we did not take into account in the discriminator or MCMC scans. 
}
\label{fig:mssm:ssc}
\end{figure}

The results can be seen in figure~\ref{fig:mssm:ssc}. 
In the MCMC scan, a large amount of points end up in the area where $m_0 \in [100, 500]$ and $m_{1/2} \in [750, 2000]$. %
This is inefficient because there are lots of would-be good points in that area, and because the rest of the parameter space remains relatively unexplored in consequence. %
In contrast, the AL scan clearly favours the regions which are on the border between good and bad. %
It nonetheless explores the rest of the parameter space in sufficient detail. %
The comparatively less explored bare band region \textit{roughly $10\%$ away from the border region} comes about as an artifact from the diversity measure which penalises points for being too close to those already chosen. 

In these figures, we have also shown a discriminator line for both scans. %
This was achieved by retraining a neural network with the same settings as the original discriminator from the AL scan on either set of points. %
To make these two lines comparable to one another, not the original discriminator but the one retrained on the whole set of points was used to produce the line for the AL scan. %
Such a retraining of networks can of course in general lead to better discriminator performance. %
This is because the gradual introduction of more interesting points, i.e. points near the boundaries, can lead to distortions of the sort where the first points, which were less interesting, have a larger impact on the network than points which follow. %
With the points distributed as they are, it is not surprising that the line from the AL points traces more accurately the boundary between good and bad points. If we define ``interesting'' points to be those near the decision boundary, so for $m_{1/2} < 250$ GeV, or all values to the left of a line from $(100,250)$ GeV to $(350, 1500)$ GeV, then the active learning scan delivered $42\%$ of its points in this region compared to only $32\%$ for the MCMC.   

In conclusion, even in a relatively simple and low-dimensional model like the CMSSM, the AL scan produces a superior choice of points than the MCMC scan. 
This has the advantage that one could use fewer points to find the boundaries, and we expect that the trained network on a set of points deliberately selected to locate the boundaries should have superior performance at locating the boundary; this can clearly be seen from the two plots. 
The cost of this, however, is that training the discriminator after each new set of points takes a certain amount of time. %
For such a simple scenario where we only take into account the dark matter density, there is no gain from using deep learning. However, if we were to include more constraints -- especially collider constraints -- we would expect that the time spent training networks to select points would be worth it. 


\section{Parameter space of the SMSQQ model}
\label{sec:smsqq}

\subsection{Highest singlet mass}
\label{sec:smsqq:msmax}

We now move to explore the SMSQQ, a model with colourful mediators, with an active learning scan. %
This model is particularly useful to illustrate how colourful unitarity bounds contribute to a mass limit for dark matter. %
In our previous paper~\cite{Goodsell:2020rfu}, we had used an iterative approach using several MCMC scans, narrowing in on the region we were interested in. %
The likelihood function used for the MCMC was artificially weighted in a way that forced it to prioritise points with a higher mass. %
Doing so, we were able to establish an upper bound for the singlet mass.

In this work, we further refined the search for a highest singlet mass by choosing new, updated ranges. %
The old and new ranges can be seen in table~\ref{tab:smsqq:ranges}. %
The results are shown in table~\ref{tab:smsqq:maxms}. %
One can see that by limiting $\Lambda$ to a high but valid range, we were able to gain almost 1 TeV on the highest singlet mass. %
In this table, we also show the points with the highest singlet mass in an AL and an MCMC scan when we only generated 200k points and used looser ranges. %
One can see that these two masses are very close together, the one produced by the AL scan even being slightly higher than the one from the MCMC. %
This is remarkable because, unlike the MCMC, the AL scan is by no means skewed towards higher masses. %

\begin{table}[h]
  \vspace{3mm}
  \footnotesize
  \centering
  \begin{tabularx}{0.88\textwidth}{|X|l|l|l|l|l|}
    \hline
    Scan & $m_S\ [\text{GeV}]$ & $m_O\ [\text{GeV}]$ & $m_E\ [\text{GeV}]$ & $\kappa\ [\text{GeV}]$ & $\Lambda$ \\
    \hline
    \hline
    old MCMC (1m points) & 47,354 & 53,875 & 39,025 & 174,121 & 3.05993 \\
    new MCMC (1m points) & \textbf{48,392} & 56,583 & 42,921 & 185,782 & 3.12349 \\
    MCMC (200k points) & 45,831 & 55,334 & 38,514 & 173,010 & 3.09748 \\
    AL (200k points) & 45,964 & 50,699 & 38,116 & 169,510 & 3.11153 \\
    \hline
  \end{tabularx}
\caption{Maximum singlet mass found in old MCMC scan and new MCMC scan with 1 million points each.
The maximum singlet masses of the 200k MCMC and AL scans are also shown.
The latter are very close to one another, despite the fact that only the MCMC was explicitly forced to prioritize high masses.
}
\label{tab:smsqq:maxms}
\end{table}

\begin{table}[h]
  \footnotesize
  \centering
  \begin{tabularx}{0.5\textwidth}{|l|l|X|}
    \hline
	  Setting & 50k scan & 200k scan \\
    \hline
    \hline
	  Number of points & 50,000 & 200,000 \\
	  Initial points & 0 & 50,000 from previous scan \\
	  K & 1000 & 300 \\
	  L & 500000 & 100000 \\
	  FullTrain & 0 & 0 \\
	  Hidden Layers & 2 & 5 \\
	  Hidden Size & 100 & 200 \\
	  Learning Rate & 0.001 & 0.0001 \\
	  SGD momentum & 0.1 & 0.1 \\
	  Training steps & 1000 & 5000 \\
	  Weight decay & 0.001 & 0.001 \\
	  Epsilon & 0.9 & 0.9 \\
	  Diversity Alpha & 0.5 & 0.5 \\
    \hline
  \end{tabularx}
\caption{Network settings for the two main AL scans in the SMSQQ.
The learning rate of the larger network is smaller because it otherwise diverges.
Explanations of each setting can be found in table~\ref{tab:toy:settings} and in the text.
}
\label{tab:smsqq:settings}
\end{table}

\begin{table}[h]
  \footnotesize
  \centering
  \begin{tabularx}{0.88\textwidth}{|X|X|X|X|X|}
    \hline
	  Variable & 50k AL scan & 200k AL / MCMC scan & max $m_S$ MCMC scan & old max $m_S$ MCMC scan~\cite{Goodsell:2020rfu} \\
    \hline
    \hline
	  $\kappa\ [\text{GeV}] $& $1.5\cdot 10^5 - 1.85\cdot 10^5$ & $5.0\cdot 10^4 - 2.0\cdot 10^5$ & $1.5\cdot 10^5 - 1.9\cdot 10^5$ & $1.5\cdot 10^4 - 1.8\cdot 10^5$ \\
	  $m_S^2 - m_E^2\ [ \text{GeV} ]^2$ & $1.0\cdot 10^8 - 1.5\cdot 10^9$ & $5.0\cdot 10^8 - 1.5\cdot 10^9$ & $1.0\cdot 10^8 - 1.5\cdot 10^9$ & $0 - 2.4\cdot 10^9$ \\
	  $m_O^2 - m_S^2\ [ \text{GeV} ]^2$ & $1.0\cdot 10^8 - 5.0\cdot 10^9$ & $1.0\cdot 10^8 - 8.0\cdot 10^9$ & $1.0\cdot 10^8 - 8.0\cdot 10^9$ & $1.0\cdot 10^4 - 2.0\cdot 10^9$ \\
	  $m_E^2\ [ \text{GeV} ]^2 $& $2.0\cdot 10^8 - 1.8\cdot 10^9$ & $1.0\cdot 10^8 - 2.0\cdot 10^9$ & $2.0\cdot 10^8 - 2.0\cdot 10^9$ & $5.0\cdot 10^4 - 2.0\cdot 10^9$ \\
	  $\Lambda$ & 2.8 - 3.2 & 0.5 - 3.4 & 2.8 - 3.2 & 0.1 - 3.7 \\
    \hline
  \end{tabularx}
\caption{Variable ranges for the main AL scans in the SMSQQ and the scans to obtain the maximum mass of the singlet.
For the new max $m_S$ scan, a tightening of $\Lambda$ proved very useful.
Tighter ranges were used for the smaller main scan to ensure that good points are found.
In the larger scan this wasn't necessary because we were able to generate good points in the vicinity of those we had already found in the previous scan.
}
\label{tab:smsqq:ranges}
\end{table}


\subsection{Performance of the AL scan}
\label{sec:smsqq:alperf}

We then move to the AL and the MCMC scan with 200k points each to explore their properties. %
The MCMC and AL scans have ranges as provided in table~\ref{tab:smsqq:ranges}. %
The settings of the AL scan can be found in table~\ref{tab:smsqq:settings}. %

We find that making an AL scan with many points is not completely straightforward though. %
Whether a network can handle a load of points depends at least in part on its number of parameters, i.e. its size. %
Other factors will be discussed in section~\ref{sec:smsqq:alvsnn}. %
An additional difficulty is that the relative scarcity of good points in the ranges we want to explore: %
In a random scan with these ranges, solely around 3/10,000 points are good. 
While an MCMC would be able to find at least \textit{one} good region somewhere despite this scarcity, the AL would not be able to operate on such few points because it does not work with gradients of any kind (as of now). %
We therefore start with a smaller AL scan with 50k points on a range of which we suspect that it contains more good points on average (see table~\ref{tab:smsqq:ranges} for these ranges). %
We then feed these 50k points, of which some 23 percent are good, to a larger network. %
This network then generates the remaining 150k points based on the lessons it has already drawn after training on the first 50k points. %
These remaining points are on the loose ranges that the MCMC scan also works on.

\begin{figure}[p]
\vspace{-30mm}
\centering
\begin{subfigure}{.45\textwidth}
  \centering
  \includegraphics[width=\linewidth]{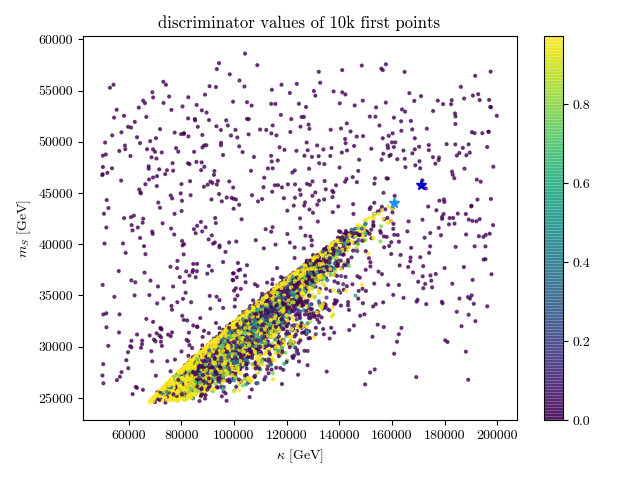}
\end{subfigure}%
\begin{subfigure}{.45\textwidth}
  \centering
  \includegraphics[width=\linewidth]{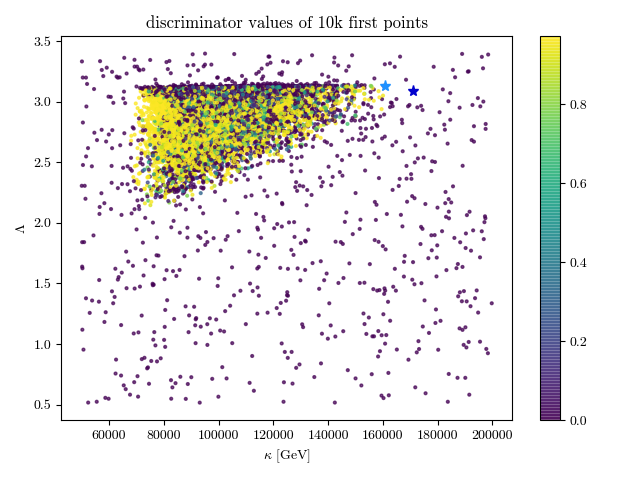}
\end{subfigure}
\begin{subfigure}{.45\textwidth}
  \centering
  \includegraphics[width=\linewidth]{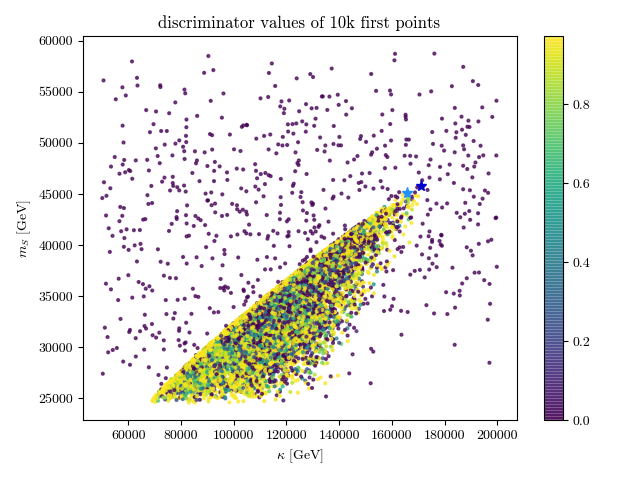}
\end{subfigure}%
\begin{subfigure}{.45\textwidth}
  \centering
  \includegraphics[width=\linewidth]{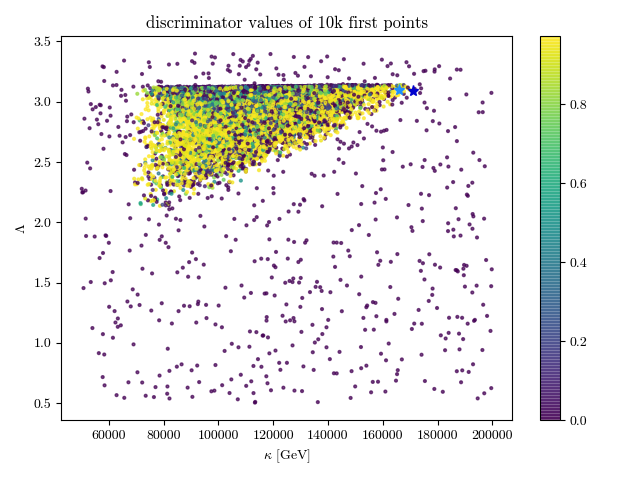}
\end{subfigure}
\begin{subfigure}{.45\textwidth}
  \centering
  \includegraphics[width=\linewidth]{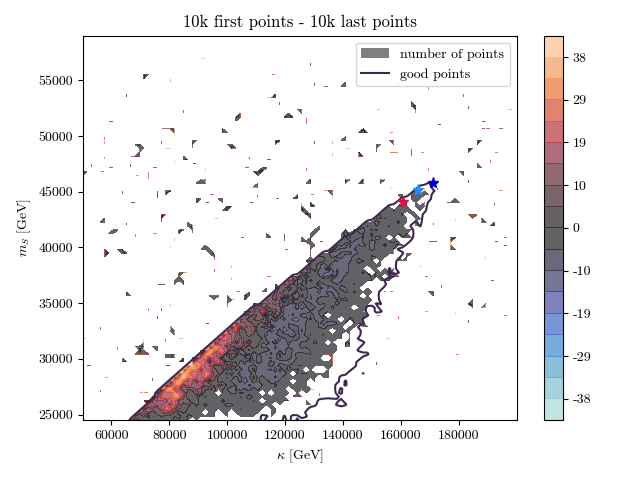}
\end{subfigure}%
\begin{subfigure}{.45\textwidth}
  \centering
  \includegraphics[width=\linewidth]{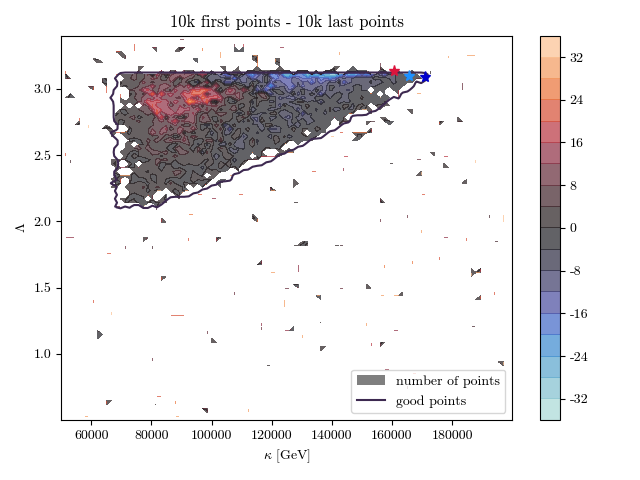}
\end{subfigure}
\caption{Comparison of the first and last 10k points in the AL SMSQQ scan with 200k points. %
Left: the $\kappa-m_S$ plane, right: the $\kappa-\Lambda$ plane. %
The uppermost plot shows the point distribution of the first 10k points, the middle one that of the last 10k points. %
In these plots, the light blue star indicates the location of the point with the highest singlet mass of that subset of data, the dark blue star the location of the highest singlet mass overall. %
The lowest plots are a comparison and summary of the two above: %
In a 100-by-100 grid, in each bin the number of points in the first and last 10k points are subtracted from one another. %
The dark blue star indicates the location of the point with the highest singlet mass $m_S$ in the whole dataset, the red one that of the first 10k points and the light blue one that of the last 10k. %
Note how, again, the area of interest broadens with time as the discriminator explores the borders of the regions of good points.
}
\label{fig:smsqq:xs-gp}
\end{figure}

\begin{figure}[h]
\centering
\begin{subfigure}{.45\textwidth}
  \centering
  \includegraphics[width=\linewidth]{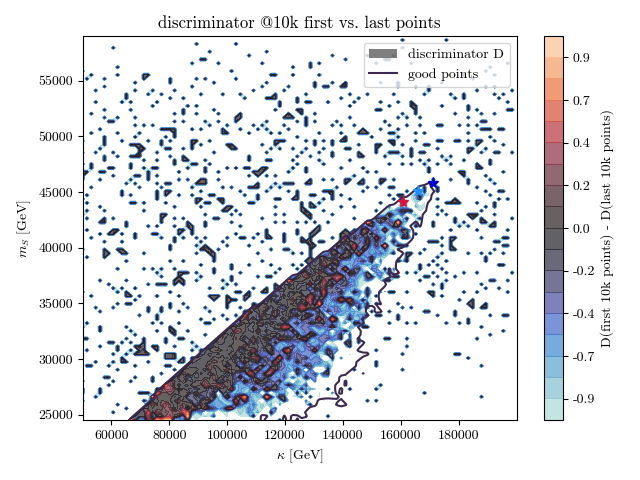}
\end{subfigure}%
\begin{subfigure}{.45\textwidth}
  \centering
  \includegraphics[width=\linewidth]{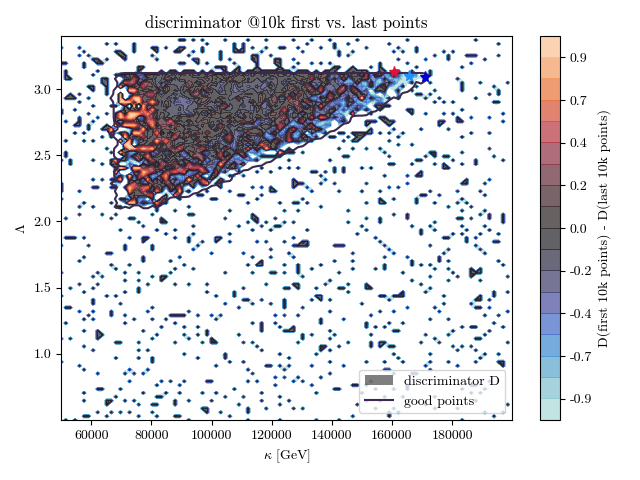}
\end{subfigure}
\caption{Comparison of the discriminator results in the first versus the last 10k AL-generated points. %
The dark blue star indicates the location of the point with the highest singlet mass $m_S$ in the whole dataset, the red one that of the first 10k points and the light blue one that of the last 10k. %
Note how the discriminator explores the border as more points get introduced. %
In each cell of a 100-by-100 bin grid, the average discriminator value of the first 10k points and the last 10k points are taken, then subtracted from one another. %
Thus, red or orange indicates areas where the discriminator was confident about finding good points among the first 10k points. %
Dark or light blue indicates areas where it was confident about finding good poitns among the last 10k. %
Grey areas indicate that the confidence in these points is roughly equal for the first and last 10k points. %
}
\label{fig:smsqq:gs}
\end{figure}

The difference between the first points and last 10k points of the AL scan can be seen in figure~\ref{fig:smsqq:xs-gp}. %
The first feature that meets the eye is the fact that the model really starts exploring the parameter space, and gets increasingly confident about areas it barely covered in the first 10k points. %
This is especially apparent in the $\kappa-m_S$ plane. %
From the summary plots in the last row of plots in the figure it is evident how much exploring the parameter space helps find a point with a higher singlet mass, too. %
We have cross-checked this phenomenon with AL scans of the same model which differed from the one shown here in various settings or point numbers.

Figure~\ref{fig:smsqq:gs} is similar to the last row in figure~\ref{fig:smsqq:xs-gp}, except that now not the point density but the average discriminator value is shown. %
One can see that there are regions which the first 10k points already investigated sufficiently (in red), and regions which the last 10k points investigated and the discriminator is quite confident about (in blue). %
Grey regions are regions where the discriminator values of the first 10k points and the last 10k points are rougly equal, or where only one of the two exists and is close to zero. %
Note that the random points outside the region are grey with a blue border; however, this is an artifact from the grid interpolation. %
This can be verified by checking top rows of the~\ref{fig:smsqq:xs-gp} plot: %
Outside the region with good points, the discriminator is always close to zero.

\begin{figure}
\centering
\begin{subfigure}{.45\textwidth}
  \centering
  \includegraphics[width=\linewidth]{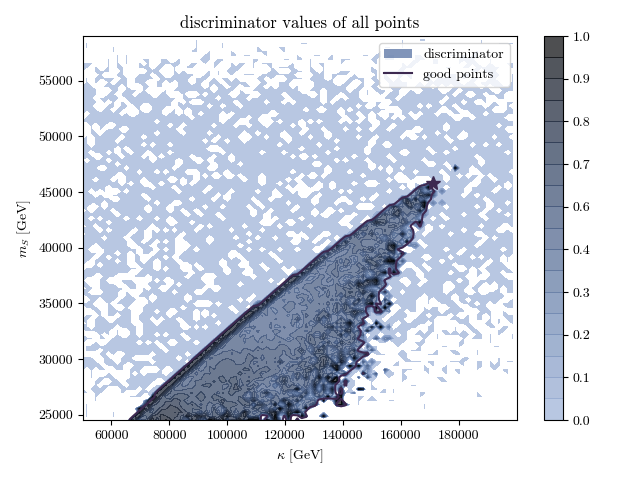}
\end{subfigure}%
\begin{subfigure}{.45\textwidth}
  \centering
  \includegraphics[width=\linewidth]{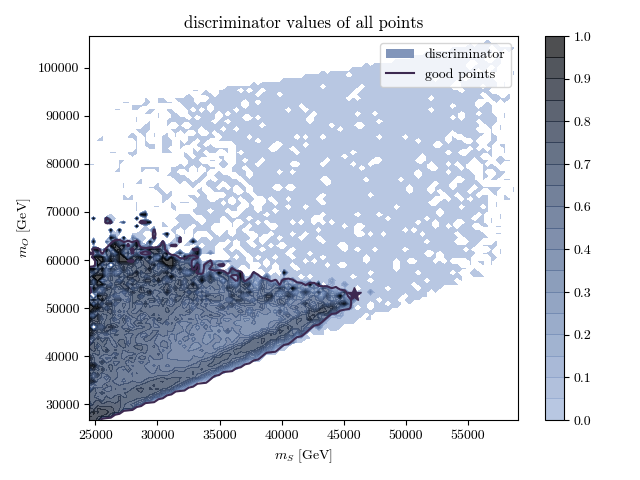}
\end{subfigure}
\begin{subfigure}{.45\textwidth}
  \centering
  \includegraphics[width=\linewidth]{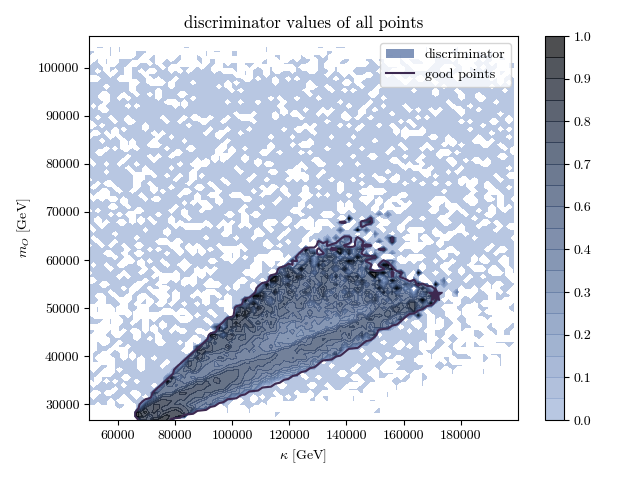}
\end{subfigure}%
\begin{subfigure}{.45\textwidth}
  \centering
  \includegraphics[width=\linewidth]{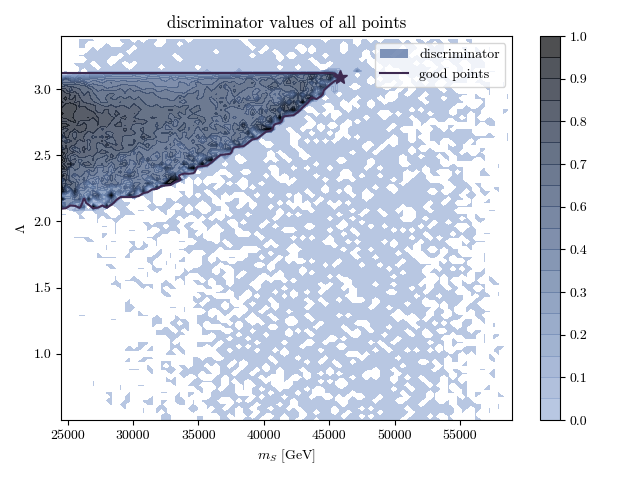}
\end{subfigure}
\caption{Distribution of the discriminator values interpolated from a 100x100 grid.
The dark purple star indicates the location of the point with the highest singlet mass $m_S$.
Take note of the substructures in the regions with good points: 
Along some borders, like e.g. the upper border in the $\kappa - m_S$ plane, the discriminator is quite sure about finding good points.
Along other borders, for example the upper border in the $m_S - \Lambda$ plane, it is not as sure.
On ``fuzzy" borders, like e.g. the upper border in the $m_S - m_O$ plane, the discriminator tends to produce blotchy results -- as one would expect.
}
\label{fig:smsqq:gc}
\end{figure}

Figure~\ref{fig:smsqq:gc} shows the discriminator values in all of the 200k points. %
One can clearly see that the discriminator only returns values significantly larger than zero inside the region with good points, with the exception of a few outliers around the fuzzy corners of these regions, and a single outlier at a high singlet mass. %
Two substructures in these distributions are an interesting byproduct: %
First, the discriminator returns around the fuzzy edges tend to be a bit blotchy. %
This could be due to the fact that a portion of points is sampled in the vicinity of already-known good points. %
Around the fuzzy edges there is a significant proportion of bad points and an uneven distribution of good points. %
This exacerbates the effect of sampling around good points. %
Second, there are almost straight lines of points that the discriminator deems good, often close to a hard border. %
This is in parts due to the fact that more points are sampled close to borders. %
It turns out, however, that these are areas where there are indeed less bad points. %
We have verified that such a stark dark discriminator line always appears in regions close to a constraint on the dark matter density\cite{Aghanim:2018eyx}. %
It is therefore conceivable that the discriminator zeroes in on dark matter constraints particularly well, thus reducing the need to generate and evaluate bad points in those areas. %
This is rendered even more plausible by the fact that dark matter constaints tend to be pretty abrupt in comparison to those imposed by unitarity\cite{Goodsell:2020rfu,Griest:1989wd,Hedri:2014mua,vonHarling:2014kha,Cahill-Rowley:2015aea,Kahlhoefer:2015bea,Baldes:2017gzw,ElHedri:2017nny,ElHedri:2018atj,Harz:2018csl,Hektor:2019ote,Kannike:2019mzk,Alanne:2020jwx,Fuks:2020tam,Espinoza:2020qyf,Espinoza:2020kut} or vacuum stability\cite{Cao:2013wqa,He:2013tla,Cheng:2018mkc,Schuessler:2007av,SchuesslerThesis,Staub:2018vux,Baker:2020vkh}. %
Gradient-agnostic discriminators like the one we are dealing with here are particularly well-suited for handling such constraints, unlike MCMCs. %
However, the implementation of a gradient of sorts into the AL scan so that it handles less abrupt constraints similarly well is left for future work.

\begin{figure}
\centering
\begin{subfigure}{.45\textwidth}
  \centering
  \includegraphics[width=\linewidth]{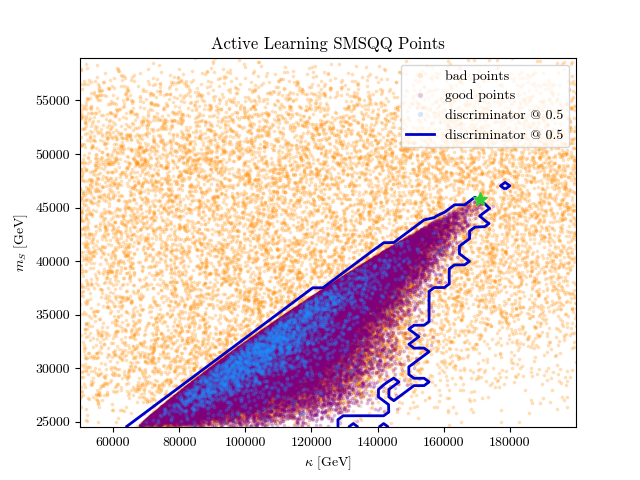}
\end{subfigure}%
\begin{subfigure}{.45\textwidth}
  \centering
  \includegraphics[width=\linewidth]{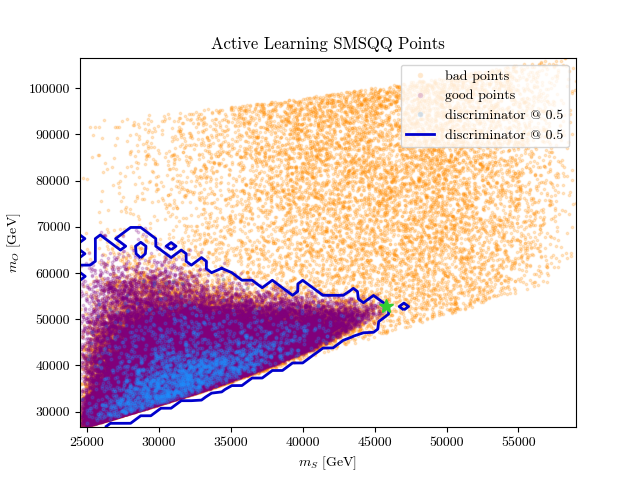}
\end{subfigure}
\caption{Scatter point distribution of good and bad AL-generated points in the SMSQQ.
The bright green point indicates the location of the point of the highest singlet mass $m_S$.
Note how the distribution of points where the discriminator is uncertain (blue points) matches the regions where there are good points and never goes to regions with bad points. 
Also note how well the discriminator identifies the borderline between good and bad points even in this higher-dimensional model, as visualized here by an interpolation grid with 100 bins on each axis.
The discriminator returns a value close to 1 where it suspects a good point, and close to 0 where it suspects a bad point.
}
\label{fig:smsqq:ssc}
\end{figure}

Figure~\ref{fig:smsqq:ssc} shows the distribution of good and bad points and the contour line of where the discriminator thinks the border between good and bad points is. %
Comparing this to the previous plots, one can see that the discriminator does a pretty good job at identifying the borderline of good points. %
Marked in blue are points where the discriminator returned a value around 0.5, i.e. where it was not sure whether this point was good or bad. %
These points are not right on the borderline as they are in our two-dimensional toy models (see figures~\ref{fig:toy:toy1} and~\ref{fig:toy:toy2} for comparison). %
This is mostly due to the higher dimensionality of this model. %
It is noteworthy as well that there is no extreme amount of blue points in the regions where the discriminator showed a large confidence for good points, as shown in figure~\ref{fig:smsqq:gc}. %
There are some points; however, especially in the $\kappa-m_S$ plane one can see that their amount is not as large as we would assume had the discriminator not been so confident around the hard upper edge of the good region. %
One can also see that the point with maximum singlet mass is right on the borderline, as was expected.


\subsection{AL vs. MCMC}
\label{sec:smsqq:alvsmcmc}

\begin{figure}
\centering
\begin{subfigure}{.45\textwidth}
  \centering
  \includegraphics[width=\linewidth]{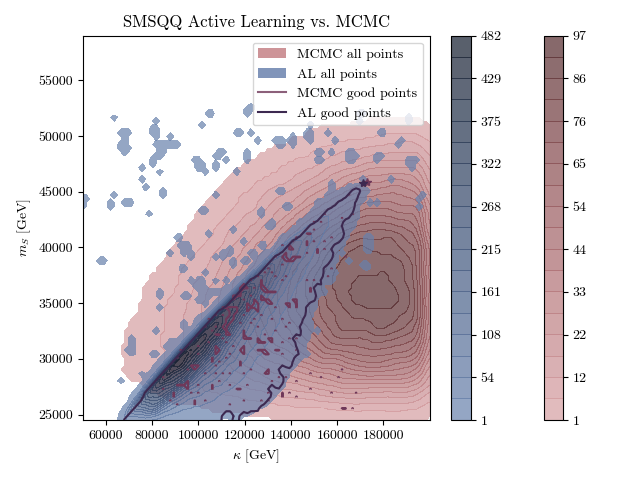}
\end{subfigure}%
\begin{subfigure}{.45\textwidth}
  \centering
  \includegraphics[width=\linewidth]{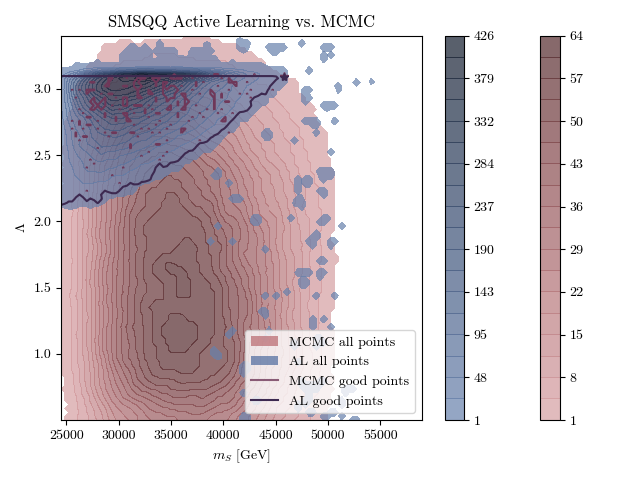}
\end{subfigure}
\begin{subfigure}{.45\textwidth}
  \centering
  \includegraphics[width=\linewidth]{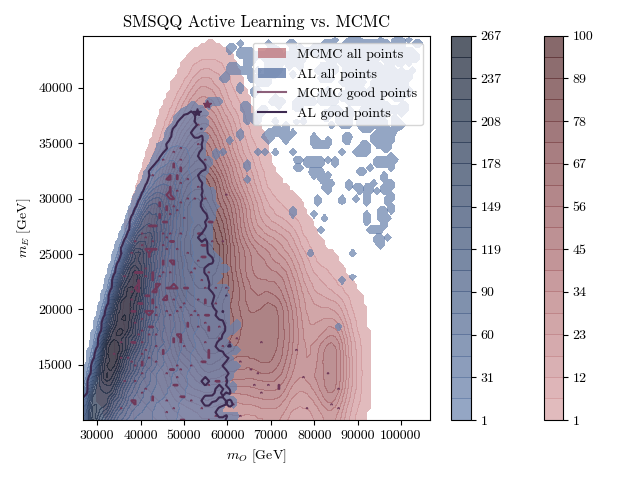}
\end{subfigure}%
\begin{subfigure}{.45\textwidth}
  \centering
  \includegraphics[width=\linewidth]{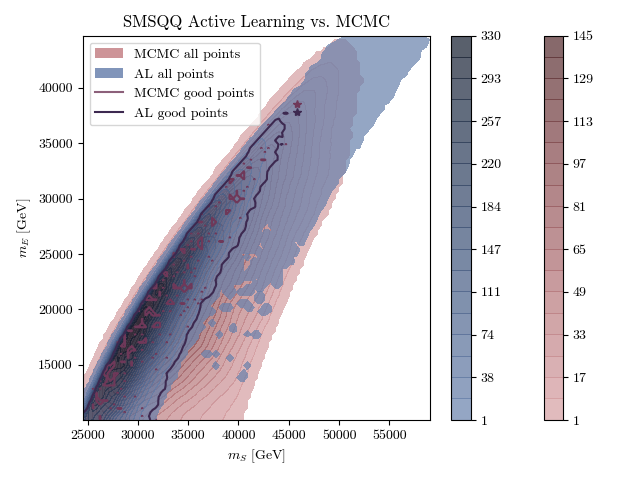}
\end{subfigure}
\caption{Comparison of MCMC- versus AL-generated points in various parameter planes of the SMSQQ model. %
The dark pink star indicates the location of the point with the highest singlet mass $m_S$ found by the MCMC, the dark blue star the one found by the AL. %
The solid lines indicate the location of good points. %
Note how many more good points the AL finds than the MCMC, and how much the region of all AL-generated points closely matches the surroundings of the good points while still exploring the parameter space.
}
\label{fig:smsqq:csx}
\end{figure}

Figure~\ref{fig:smsqq:csx} shows a comparison of the points explored by the AL versus those of the MCMC. %
The first thing to jump to the eye is how much larger the region is that the MCMC covers. %
This does not speak in favor of the MCMC, however. %
While it does find 283 good points out of 200k total (versus 70 out of 200k for a random scan), the AL finds a whopping 85,291 good points. %
Not only is this an enormous difference in numbers; as we established earlier, the good points that the AL scan finds tend to be of a higher quality because they are more often than not close to the borders of the good regions and thus help finding boundaries better. %
On the other hand, the AL scan failed to find a region of good points at $\kappa > 140,000$ GeV and $m_S <30,000$ GeV. %
This failure might have been mitigated by adding more random points to explore the rest of the parameter space, increasing the diversity measure, or generating more initial random points. %
Nevertheless, this shows that the AL scans can be quite sensitive to tuning parameters. %
We deem it feasible to build a similar scan which tunes relevant parameters automatically; %
however, we leave this for future work.

On the whole, and despite the one non-identified region, the advantages of AL scans are quite apparent from this figure.
In higher-dimensional parameter spaces where good points are scarce, boundaries are difficult to predict, and points take a long time to evaluate with established HEP tools, AL scans propose a much more efficient way to explore parameter spaces.


\subsection{AL vs. other networks}
\label{sec:smsqq:alvsnn}

\begin{table}[h]
\centering
\begin{tabularx}{0.88\textwidth}{|l|l|X|X|X|}
  \hline
  \multirow{2}{*}{classifier} & \multirow{2}{*}{training points} & \multicolumn{3}{c|}{percent error testing on} \\
                                                                 \cline{3-5}
                              &                                  & 200k random points & 200k AL points & 24k RFC-misclassified AL points \\
  \hline
  \multirow{2}{*}{RFC} & AL 200k & 18.7 & 27.2 & {\color{gray}100} \\
                       \cline{2-5}
                       & Random 200k & 4.8 & 44.0 & {\color{gray}100} \\
  \hline
  \multirow{2}{*}{AL} & AL 200k & 1.6 & 13.9 & 21.6 \\
                      \cline{2-5}
                      & AL 50k & 1.6 & 28.4 & 38.5 \\
  \hline
  \multirow{4}{*}{NN} & AL 200k & -- & -- & -- \\
                      \cline{2-5}
                      & AL 50k & 1.6 & 28.4 & 38.5 \\
                      \cline{2-5}
                      & Random 200k & -- & -- & -- \\
                      \cline{2-5}
                      & Random 50k & 15.8 & 42.9 & 49.1 \\
  \hline
\end{tabularx}
\caption{Benchmark comparison of RFC, AL, and various neural networks (NN). %
Each classifier was trained on one set of either 50k or 200k points which are either AL-generated or random.
The 50k points that serve the AL as an initial dataset already lead to a very good result; therefore the active learning and the simple neural network are identical in this case.
The neural network fails to train on 200k points; it stays at a 50\% error throughout the process.
The 24k RFC-misclassified points come from the RFC trained on 50k active learning points and tested on 50k active learning points from a separate run.
Note that the amount of RFC-misclassified points tallies up with the error rate of 44.0\% on 200k test points because, after rebalancing the test data set to contain equal amounts of good and bad points, we're left with about 24k test points.
}
\label{tab:smsqq:benchmark1}
\end{table}

\begin{table}[h]
\centering
\begin{tabularx}{0.88\textwidth}{|l|l|X|X|X|}
  \hline
  \multirow{2}{*}{classifier} & \multirow{2}{*}{training points} & \multicolumn{3}{c|}{percent error testing on} \\
                                                                 \cline{3-5}
                              &                                  & 50k random points & 50k AL points & 12k RFC-misclassified AL points \\
  \hline
  \multirow{2}{*}{RFC} & AL 50k & 33.6 & 31.1 & {\color{gray}100} \\
                       \cline{2-5}
                       & Random 50k & 6.2 & 44.7 & {\color{gray}100} \\
  \hline
  \multirow{2}{*}{AL} & AL 50k & 29.4 & 24.7 & 31.3 \\
                      \cline{2-5}
                      & AL 13k & 16.8 & 30.6 & 54.4 \\
  \hline
  \multirow{4}{*}{NN} & AL 50k & -- & -- & -- \\
                      \cline{2-5}
                      & AL 13k & 19.6 & 13.1 & 19.2 \\
                      \cline{2-5}
                      & Random 50k & -- & -- & -- \\
                      \cline{2-5}
                      & Random 13k & 30.6 & 40.0 & 42.8 \\
  \hline
\end{tabularx}
\caption{Benchmark like before but on the first 50k points, with smaller ranges and only K = 500 initial points.
The neural network fails to train properly with 50k points, but succeeds with the smaller datasets.
}
\label{tab:smsqq:benchmark2}
\end{table}

In a final step, we benchmark how well our AL models do against other networks in the SMSQQ. %
We set out generating two new AL scans with 200k points each, but this time stopping the training only after the error rate has dropped below 0.5 percent (compared to 5 percent earlier). %
We then train two neural networks (NN) and a random forest classifier (RFC) with the exact same settings on these points and a set of random points. %
We also try to find the amount of points after which the error rate falls below 20 percent; %
however, this is already reached after 50k points because the network is larger than the one used to generate these 50k initial training points. %
It is for this reason that the neural network and the active learning scan, trained on 50k points, do identically well (see table~\ref{tab:smsqq:benchmark1}). 

We also see that the error rates of the RFC, kept on its original settings of 150 estimators, are vastly worse than those of the AL and NN. %
It is worth mentioning that the NN \textit{fails to train} on 200k points, regardless of whether they are random or AL-generated. %
This cannot be solely because of the amount of points, as the AL trains just fine on 200k points. %
Rather, it either fails because the network parameters are not ideal for getting all points at once, or it fails because being presented with all these points at once is fundamentally too much for a network this size. %
As is often the case in this area of research, it is difficult to prove the one or the other, though. %
The AL network does fine, however, by getting the points in a piecemeal fashion, which gives it the chance to adjust its weights gradually. 

As expected, the NN trained on 50k random points does consistently worse than that trained on 50k AL-generated points because the quality of AL-generated points is higher in the sense that there are more points in areas where there is a lot to learn. %
There is no difference between the AL trained on 50k and 200k points when tested only on random points, differences appear when testing with another set of AL-generated points (which are presumably more interesting). %
This difference further manifests when tested on only those points which are AL-generated and an RFC, trained on AL-generated points, misclassified (these points are potentially even more interesting). 

Keeping this in mind, we move to make two more AL scans of 50k points and a smaller network, with settings like the previous AL 50k scan (see table~\ref{tab:smsqq:settings}). %
We find that the error rate of the AL network drops below 20 percent after about 13k points when tested on a set of 200k random points. %
As before, the NN fails to train with this amount of points. %
Unsurprisingly, the NN trained on 13k random points does fairly poorly, especially on interesting points. %
Surprisingly, though, the NN trained on 13k AL-generated points does better than the AL model on these points, particularly for interesting points. %
This might be happening due to the fact that there are no initial points in the AL, meaning that the first round of K points might be getting an overly large amount of attention. %
There is a fairly easy fix to this, though: adding a sufficiently large amount of random points as initial training set into the AL should do the trick, like it was necessary for some of the toy models in section~\ref{sec:toy}. %
It is somewhat surprising that the RFC does comparably well when it is trained and tested on random points. %
This seeming superiority vanishes quickly, however, when tested on more interesting points. %
As before, the performance of the AL improves with more points. %
This is not a monolith, however: The AL trained on 50k points actually does \textit{worse} than that trained on 13k points, when tested on random points. %
It is better, however, at judging more interesting points. 

Overall, this benchmarking effort demonstrates two advantages that can be generalised beyond the confines of this particular model:
First, AL scans are very good at finding interesting points and at scoring well when tested on similarly interesting points. %
This makes them good for finding boundaries in new models. %
Second, by using the fact that AL scans feed points to their network in a piecemeal fashion, we are able to get away with smaller networks than we would if we somehow found interesting points and subsequently trained a network on these. %
These two advantages make AL scans particularly useful for working on new, complex, and higher-dimensional models in a cost- and compute-efficient way. %
Nevertheless, there is one obvious limitation to AL scans because many parameters need to be tuned, which requires some know-how. %
As mentioned earlier, writing an autotuning AL scan is left for future work.


\section{Learning the Higgs mass in the MDGSSM}
\label{sec:mdgssm}

\subsection{Setup and performance of the AL scan}
\label{sec:mdgssm:setup}

\begin{table}[h]
  \footnotesize
  \centering
  \begin{tabularx}{0.5\textwidth}{|l|l|X|}
    \hline
    Setting & 20k scan & 100k scan \\
    \hline
    \hline
    Number of points & 20,000 & 100,000 \\
    Initial points & 0 & 20,000 from previous scan \\
    K & 100 & 500 \\
    L & 50000 & 50000 \\
    FullTrain & 0 & 0 \\
    Hidden Layers & 2 & 5 \\
    Hidden Size & 100 & 200 \\
    Learning Rate & 0.001 & 0.0001 \\
    SGD momentum & 0.1 & 0.05 \\
    Training steps & 200 & 5000 \\
    Weight decay & 0.001 & 0.001 \\
    Epsilon & 1 & 0.8 \\
    Diversity Alpha & 0.1 & 0.1 \\
    \hline
  \end{tabularx}
\caption{Network settings for the two AL scans in the MDGSSM.
As in the SMSQQ, the learning rate of the larger network is smaller, otherwise it diverges.
Explanations of the settings can be found in table~\ref{tab:toy:settings} and in the text.
}
\label{tab:mdgssm:settings}
\end{table}

\begin{table}[h]
  \footnotesize
  \centering
  \begin{tabularx}{0.3\textwidth}{|X|X|}
    \hline
    Variable & Range \\
    \hline
    \hline
    $m_{DY}\ [\text{GeV}] $& 56 - 2016 \\
    $m_{D2}\ [ \text{GeV} ]$ & 199 - 1532 \\
    $\mu\ [ \text{GeV} ]^2$ & 202 - 2056 \\
    $\tan\ \beta $& 5.45 - 34.95 \\
    $-\lambda_S$ & 0.01 - 1.46 \\
    $\sqrt{2}\lambda_T$ & -1.39 - 0.44 \\
    \hline
  \end{tabularx}
\caption{Variable ranges for the two AL scans in the MDGSSM.
The ranges were chosen so that they include all 10 reference points in~\cite{Goodsell:2020lpx}.
}
\label{tab:mdgssm:ranges}
\end{table}

\begin{figure}
\centering
\begin{subfigure}{.45\textwidth}
  \centering
  \includegraphics[width=\linewidth]{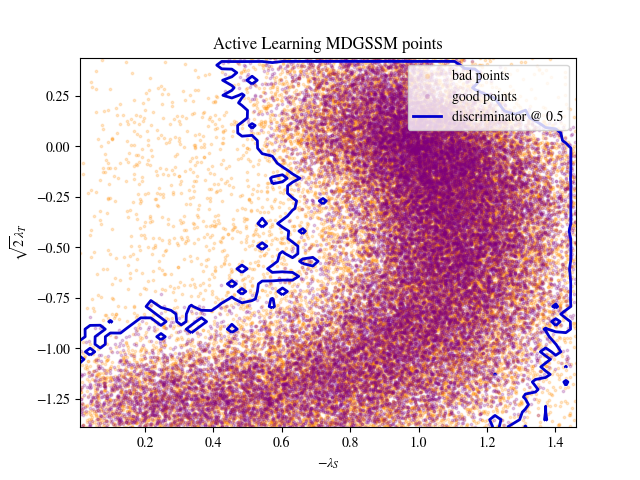}
\end{subfigure}%
\begin{subfigure}{.45\textwidth}
  \centering
  \includegraphics[width=\linewidth]{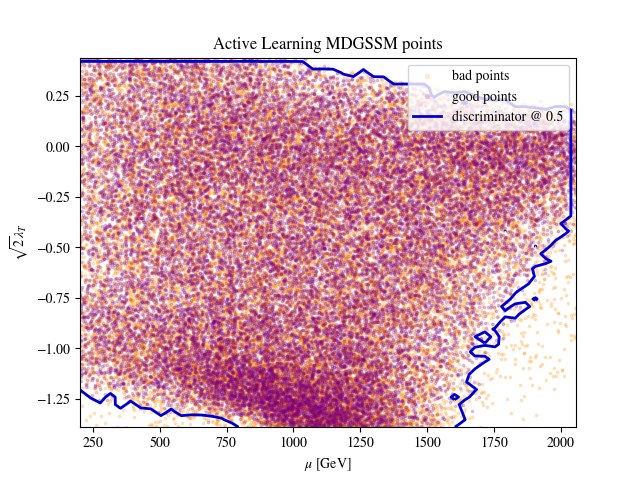}
\end{subfigure}
\begin{subfigure}{.45\textwidth}
  \centering
  \includegraphics[width=\linewidth]{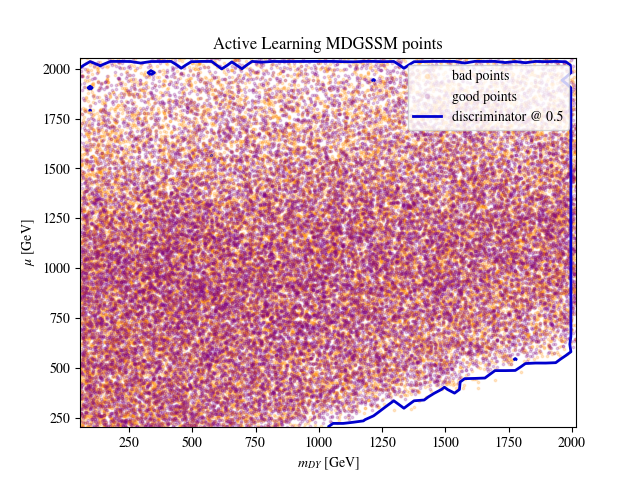}
\end{subfigure}%
\begin{subfigure}{.45\textwidth}
  \centering
  \includegraphics[width=\linewidth]{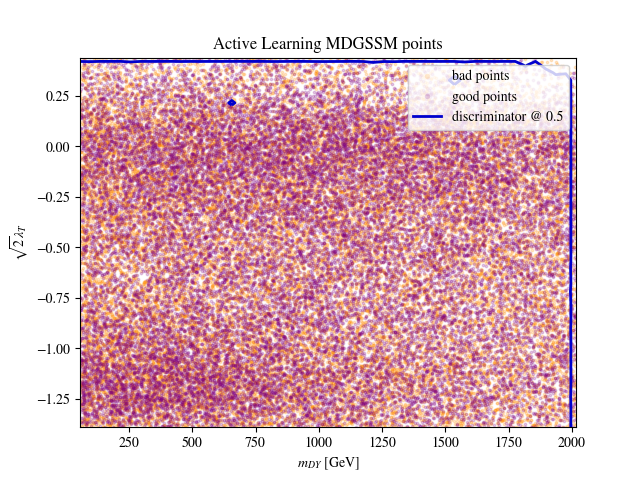}
\end{subfigure}
\caption{Distribution of good and bad points in various planes of the MDGSSM.
Generally speaking there is visible structure for planes along $-\lambda_S$, and just a little to almost no structure in all other planes.
Even in the more insightful plots, it is hard to visually separate good regions from bad ones.
}
\label{fig:mdgssm:ssc}
\end{figure}

In a final step, we apply the AL scan to the Minimal Dirac Gaugino Supersymmetric Standard Model (MDGSSM). %
This is a non-minimal supersymmetric scenario with many parameters at low energy. Collider constraints on strongly-coupled particles on this model were considered in \cite{Chalons:2018gez}. Subsequently in \cite{Goodsell:2020lpx} the constraints on electroweak-charged particles were considered from both dark matter and collider searches. As a consequence of the constraints on colourful particles, it is prudent to consider them heavy (of order $2$ to $3$ TeV) where their exact values do not significantly affect the phenomenology. This leaves six interesting parameters for the low energy theory that greatly affect the masses and phenomenology of the electroweak sector. To place constraints, it is therefore necessary to scan over these parameters. However, it was found in \cite{Goodsell:2020lpx} that only a small proportion of parameter choices lead to an acceptable Higgs mass. Therefore a random forest classifier was first trained on a random dataset and used to filter proposed points in a larger MCMC scan over the dark matter parameter space.


Here we shall examine whether AL can do a better job at this task, by selecting points to train a discriminator that decides whether a given point has a Higgs mass in the range 122 - 128 GeV. In particular, the \SARAH code for the MDGSSM produces a lot of null points, where the result is not a bad Higgs mass but rather \texttt{NaN}. %
In contrast to a continuous fit, a discriminator like the one an AL scan employs is naturally equipped to deal with such outcomes. %

We proceed in a two-step fashion again, putting a smaller network to work on 20k points, then feeding those points to a larger network and generating 100k points in total. %
(We are generating less points than for the SMSQQ for the simple reason that points in the MDGSSM take longer to generate.)
As before, the larger network requires a smaller learning rate; in this model, a smaller stochastic gradient descent momentum stabilizes the larger network as well. %
The ranges of this scan are chosen such that they include all reference points listed in the literature~\cite{Goodsell:2020lpx}. %
They are listed in table~\ref{tab:mdgssm:ranges}. %
Note that we need not make tighter ranges for the smaller scan in this model because there is a sufficient amount of good points in the chosen ranges (around 7 percent of randomly selected points are good). %
Nevertheless, the two-step approach is justified because, as we saw in section~\ref{sec:smsqq:alvsnn}, feeding a generous amount of initial training points from the first run to the larger network brings better training results.

Figure~\ref{fig:mdgssm:ssc} shows the result of this scan in various planes. %
These distributions do not at all exhibit neat, spatially secluded regions of good points like we saw with the SMSQQ in the previous section. %
Instead, good and bad points are pretty much all jumbled together, with structures visible mostly in planes containing $\sqrt{2}\lambda_T$ and very little structure otherwise. %
This makes it hard to separate good regions from bad ones in a spatial way, as illustrated by the large areas that the blue line in the plots encompasses. %
This occurs not as a fault of the scan, but as an intrinsic feature of the model: %
Several variables in the MDGSSM only have an indirect effect on the Higgs mass, such that their influence does not show in a plot. %
Crucially, however, a neural network -- or an AL scan for that matter -- is still able to use these variables as a basis to estimate whether they give rise to a good or bad Higgs mass. %
This is demonstrated in the section below.


\subsection{AL vs. other networks}
\label{sec:mdgssm:alvsnn}

\begin{table}[h]
\centering
\begin{tabularx}{0.88\textwidth}{|l|l|X|X|X|}
  \hline
  \multirow{2}{*}{classifier} & \multirow{2}{*}{training points} & \multicolumn{3}{c|}{percent error testing on} \\
	                                                         \cline{3-5}
			      &  			 	 & 62k random points & 63k AL points & 17k RFC-misclassified AL points \\
  \hline
  \multirow{2}{*}{RFC} & AL 58k & 18.2 & 31.6 & \textcolor{gray}{100} \\
		       \cline{2-5}
		       & Random 62k & 19.3 & 32.6 & \textcolor{gray}{100} \\
  \hline
  \multirow{2}{*}{AL} & AL 58k & 3.3 & 10.4 & 13.8 \\
	              \cline{2-5}
		      & AL 24k & 6.7 & 17.3 & 22.3 \\
  \hline
  \multirow{4}{*}{NN} & AL 58k & 2.9 & 5.6 & 9.3 \\
	              \cline{2-5}
		      & AL 24k & 5.3 & 14.1 & 23.0 \\
		      \cline{2-5}
		      & Random 62k & 2.4 & 7.9 & 13.7 \\
		      \cline{2-5}
                      & Random 24k & 5.9 & 16.4 & 23.8 \\
  \hline
\end{tabularx}
\caption{Benchmark of AL, RFC, and neural networks on the MDGSSM. %
There are 58k-63k non-null AL-generated points out of 100k total points, and 62k random points of 100k. %
The 14k points were selected based on the number of points after which the original AL (with 58k points total) reached an error below 5\%. %
The fact that the 14k AL does not do as well is due to statistical fluctuations.
One can see that while the RFC trails far behind, the AL and NN do similarly well on the AL-generated points.
The NNs trained on random points yield similar results but become less reliable with RFC-misclassified points when the number of training points is large.
}
\label{tab:mdgssm:benchmark}
\end{table}

In the same fashion as with the SMSQQ, we now benchmark our AL scan against various neural networks and RFCs. %
The results are shown in table~\ref{tab:mdgssm:benchmark}. %
Regularly testing on a dataset of 20k random points, we find that the error rate drops below 5 percent after around 24k not-null points, i.e. after adding about 7 sets of K new points to the initial dataset of 20k points. %
This is a substantially better error rate than the one we get with the SMSQQ. %
We should not overinterpret this, however, because the distributions of good and bad points are vastly different from those in the SMSQQ. %
Note that the AL scan that was retrained on another 24k points scored slightly more than 5 percent, which is due to statistics. %
This tells us that there is some uncertainty to all values shown in this table. %
Nevertheless, they give us an idea of where we are going with this.

As expected, the AL trained on 58k points outperforms the one that trained on 24k points only. %
This means that new interesting points add value to the overall performance of the network. %
The neural networks trained on the AL-generated points do similarly well as the AL networks. %
In the case of the 58k training points, the NN slightly outperforms the corresponding AL network. %
This highlights the importance of training the AL on its full dataset from time to time; this was not done with this model due to the fact that it requires further finetuning of the network to ensure stability in subsequent training rounds. %
Regularly retraining the AL network on its full dataset therefore is, as of now, only recommended for simpler models -- we did this for the toy models in section~\ref{sec:toy} but not in these more advanced ones. %
It is conceivable, however, that this might be more feasible with an autotuning AL scan, which we have left for future work. 

Unlike the results for the SMSQQ (see tables~\ref{tab:smsqq:benchmark1} and~\ref{tab:smsqq:benchmark2}), a neural network trained on random points does similarly well to the one trained on AL points. %
Although all networks shown here handle the impact of indirectly influencing variables well, this shows a limitation to AL scans: %
when regions of good points are not clearly distinguishable from regions of bad points, the value of generating interesting points is of limited value. %
It is possible that a set of random points, fed to a sufficiently large neural network, will do a similarly good job as an AL scan in the role of a ``Higgs-gatekeeper.'' %
There are two reasons why we would advise using an AL scan nevertheless: %
First, it is not always clear which spatially separated structures the AL might uncover anyways, which would allow it to generate at least \textit{some} interesting points. %
As one can see from the fact that the error rates of tests on random points and AL-generated points do differ in a statistically significant way, the AL scan has managed to produce at least somewhat interesting points even with this model. %
Second, as we have seen in section~\ref{sec:smsqq:alvsnn}, AL scans might allow us to get away with smaller networks by feeding it piecemeal portions of points at a time. %
Both reasons imply that AL scans are a more computational- and cost-efficient procedure. %

In summary, even in fairly complex models like the MDGSSM, AL scans are a cost-efficient way of producing gatekeepers, be it for the Higgs mass or for another observable. %
This helps us save compute resources which, in other scans, might have gotten wasted using very effective but nevertheless time-consuming HEP tools. %
Feeding the ``Higgs-gatekeeper'' we presented here to an MCMC- or other scan is left for future work. %
The standout feature we see from the performance analysis on the MDGSSM is that the concept of such gatekeepers from AL scans is quite generalizable and can be used for all kinds of models.


\section{Conclusions}
\label{sec:conc}

In this work we propose a novel approach to explore the parameter spaces of new models. %
We have demonstrated on a variety of different models that active learning scans provide a cost- and computational-efficient way to find boundaries and identify areas with good points. %
We have also shown the use of this approach even with models that do not have a spatially secluded region of good points, as is the case in the MDGSSM. %
In such cases, AL scans can be used to produce so-called gatekeepers, i.e. networks which are able to predict whether the observables resulting from a set of variables are good or bad with relative certainty. %
These can be plugged into subsequent MCMC- or other scans to avoid spending unnecessary time and resources evaluating HEP tools on points that are most likely bad anyway.

In comparison to MCMC scans, AL scans do a much better job at finding and identifying regions of good points -- exactly as we advertised. %
If the initial training set is not large enough or does not cover the entire parameter space, however, there is a risk of missing potential good regions (see section~\ref{sec:smsqq:alvsmcmc}). %
To further ensure finding all good regions, one can finetune AL parameters such as the diversity measure or the proportion of random points in each training set. %
This shows the importance (and tediousness) of finetuning at this point in time. %
We think that implementing an AL scan that automatically tunes its own parameters without the need of the user's intervention is feasible; %
however, we leave this task for future work.

In comparison to RFCs, AL scans vastly outperform in terms of accuracy. %
Vanilla neural networks can sometimes slightly outperform AL networks when trained on the same points (see sections~\ref{sec:smsqq:alvsnn} and~\ref{sec:mdgssm:alvsnn}); %
however, such NNs cannot generate interesting points on their own. %
To combine the advantages of the two, one could retrain the AL network on its full dataset every so often. %
At this point in time, this is only feasible for small or simple models at this point, though, because retraining on larger models requires more finetuning of the network parameters. %
We also find that regular NNs sometimes fail to train with a large amount of training points (see section~\ref{sec:smsqq:alvsnn}). %
This might be due to suboptimal settings to train with such large datasets, or it could be that the network is fundamentally not able to cope with such many points at once. %
If it is a matter of network parameters, an autotuning AL scan might fix this problem in the future. %
One last drawback of AL scans is that we need a sufficiently large amount of good points in order to start training(see section~\ref{sec:smsqq:alperf}. %
This could be reduced in the future by introducing gradients to the selection of K from L training points. %
This, too, is left for future work.

Finally, the code for this work will be released as part of a general framework for running simple scans along with a future publication. At that point it would also be interesting to consider more sophisticated deep learning and AL approaches such as those in the {\tt swyft} library\cite{Miller:2020hua}, and whether they can be used to improve the performance of tasks that we are interested in here. 



\section*{Acknowledgements}

MDG acknowledges support from the grants
\mbox{``HiggsAutomator''} and \mbox{``DMwithLLPatLHC''} of the Agence Nationale de la Recherche
(ANR) (ANR-15-CE31-0002), (ANR-21-CE31-0013). We thank Humberto Reyes and Sabine Kraml for collaboration on related topics and helpful discussions. 

\bibliographystyle{h-physrev}
\bibliography{lit}

\begin{thebibliography}{10}

\bibitem{Slavich:2020zjv}
P.~Slavich, \emph{et~al.},
\newblock ``{Higgs-mass predictions in the MSSM and beyond}'',
\newblock \href{http://dx.doi.org/10.1140/epjc/s10052-021-09198-2}{Eur. Phys.
  J. C\,\textbf{81},\,450\,(2021)},
  \href{http://arxiv.org/abs/2012.15629}{arXiv:2012.15629}.

\bibitem{Staub:2008uz}
F.~Staub,
\newblock ``{SARAH}'',
\newblock \,\,(2008), \href{http://arxiv.org/abs/0806.0538}{arXiv:0806.0538}.

\bibitem{Staub:2013tta}
F.~Staub,
\newblock ``{SARAH 4 : A tool for (not only SUSY) model builders}'',
\newblock \href{http://dx.doi.org/10.1016/j.cpc.2014.02.018}{Comput. Phys.
  Commun.\,\textbf{185},\,1773\,(2014)},
  \href{http://arxiv.org/abs/1309.7223}{arXiv:1309.7223}.

\bibitem{Goodsell:2015ira}
M.~Goodsell, K.~Nickel, F.~Staub,
\newblock ``{Generic two-loop Higgs mass calculation from a diagrammatic
  approach}'',
\newblock \href{http://dx.doi.org/10.1140/epjc/s10052-015-3494-6}{Eur. Phys. J.
  C\,\textbf{75},\,290\,(2015)},
  \href{http://arxiv.org/abs/1503.03098}{arXiv:1503.03098}.

\bibitem{Goodsell:2015yca}
M.~D. Goodsell, K.~Nickel, F.~Staub,
\newblock ``{The Higgs Mass in the MSSM at two-loop order beyond minimal
  flavour violation}'',
\newblock \href{http://dx.doi.org/10.1016/j.physletb.2016.04.034}{Phys. Lett.
  B\,\textbf{758},\,18\,(2016)},
  \href{http://arxiv.org/abs/1511.01904}{arXiv:1511.01904}.

\bibitem{Braathen:2017izn}
J.~Braathen, M.~D. Goodsell, F.~Staub,
\newblock ``{Supersymmetric and non-supersymmetric models without catastrophic
  Goldstone bosons}'',
\newblock \href{http://dx.doi.org/10.1140/epjc/s10052-017-5303-x}{Eur. Phys. J.
  C\,\textbf{77},\,757\,(2017)},
  \href{http://arxiv.org/abs/1706.05372}{arXiv:1706.05372}.

\bibitem{Goodsell:2018tti}
M.~D. Goodsell, F.~Staub,
\newblock ``{Unitarity constraints on general scalar couplings with SARAH}'',
\newblock \href{http://dx.doi.org/10.1140/epjc/s10052-018-6127-z}{Eur. Phys.
  J.\,\textbf{C78},\,649\,(2018)},
  \href{http://arxiv.org/abs/1805.07306}{arXiv:1805.07306}.

\bibitem{Goodsell:2020rfu}
M.~D. Goodsell, R.~Moutafis,
\newblock ``{How heavy can dark matter be? Constraining colourful unitarity
  with SARAH}'',
\newblock \href{http://dx.doi.org/10.1140/epjc/s10052-021-09597-5}{Eur. Phys.
  J. C\,\textbf{81},\,808\,(2021)},
  \href{http://arxiv.org/abs/2012.09022}{arXiv:2012.09022}.

\bibitem{Porod:2003um}
W.~Porod,
\newblock ``{SPheno, a program for calculating supersymmetric spectra, SUSY
  particle decays and SUSY particle production at e+ e- colliders}'',
\newblock
  \href{http://dx.doi.org/10.1016/S0010-4655(03)00222-4}{Comput.Phys.Commun.\,\textbf{153},\,275\,(2003)},
  \href{http://arxiv.org/abs/hep-ph/0301101}{arXiv:hep-ph/0301101}.

\bibitem{Porod:2011nf}
W.~Porod, F.~Staub,
\newblock ``{SPheno 3.1: Extensions including flavour, CP-phases and models
  beyond the MSSM}'',
\newblock \,\,(2011), \href{http://arxiv.org/abs/1104.1573}{arXiv:1104.1573}.

\bibitem{Belanger:2018ccd}
G.~B\'elanger, F.~Boudjema, A.~Goudelis, A.~Pukhov, B.~Zaldivar,
\newblock ``{micrOMEGAs5.0 : Freeze-in}'',
\newblock \href{http://dx.doi.org/10.1016/j.cpc.2018.04.027}{Comput. Phys.
  Commun.\,\textbf{231},\,173\,(2018)},
  \href{http://arxiv.org/abs/1801.03509}{arXiv:1801.03509}.

\bibitem{Belanger:2020gnr}
G.~Belanger, A.~Mjallal, A.~Pukhov,
\newblock ``{Recasting direct detection limits within micrOMEGAs and
  implication for non-standard Dark Matter scenarios}'',
\newblock \,\,(2020), \href{http://arxiv.org/abs/2003.08621}{arXiv:2003.08621}.

\bibitem{Bechtle:2013xfa}
P.~Bechtle, S.~Heinemeyer, O.~St\r{a}l, T.~Stefaniak, G.~Weiglein,
\newblock ``{$HiggsSignals$: Confronting arbitrary Higgs sectors with
  measurements at the Tevatron and the LHC}'',
\newblock \href{http://dx.doi.org/10.1140/epjc/s10052-013-2711-4}{Eur. Phys. J.
  C\,\textbf{74},\,2711\,(2014)},
  \href{http://arxiv.org/abs/1305.1933}{arXiv:1305.1933}.

\bibitem{Bechtle:2020uwn}
P.~Bechtle, S.~Heinemeyer, T.~Klingl, T.~Stefaniak, G.~Weiglein, J.~Wittbrodt,
\newblock ``{HiggsSignals-2: Probing new physics with precision Higgs
  measurements in the LHC 13 TeV era}'',
\newblock \href{http://dx.doi.org/10.1140/epjc/s10052-021-08942-y}{Eur. Phys.
  J. C\,\textbf{81},\,145\,(2021)},
  \href{http://arxiv.org/abs/2012.09197}{arXiv:2012.09197}.

\bibitem{Bechtle:2008jh}
P.~Bechtle, O.~Brein, S.~Heinemeyer, G.~Weiglein, K.~E. Williams,
\newblock ``{HiggsBounds: Confronting Arbitrary Higgs Sectors with Exclusion
  Bounds from LEP and the Tevatron}'',
\newblock \href{http://dx.doi.org/10.1016/j.cpc.2009.09.003}{Comput. Phys.
  Commun.\,\textbf{181},\,138\,(2010)},
  \href{http://arxiv.org/abs/0811.4169}{arXiv:0811.4169}.

\bibitem{Bechtle:2020pkv}
P.~Bechtle, D.~Dercks, S.~Heinemeyer, T.~Klingl, T.~Stefaniak, G.~Weiglein,
  J.~Wittbrodt,
\newblock ``{HiggsBounds-5: Testing Higgs Sectors in the LHC 13 TeV Era}'',
\newblock \href{http://dx.doi.org/10.1140/epjc/s10052-020-08557-9}{Eur. Phys.
  J. C\,\textbf{80},\,1211\,(2020)},
  \href{http://arxiv.org/abs/2006.06007}{arXiv:2006.06007}.

\bibitem{Camargo-Molina:2013qva}
J.~E. Camargo-Molina, B.~O'Leary, W.~Porod, F.~Staub,
\newblock ``{$\mathbf{Vevacious}$: A Tool For Finding The Global Minima Of
  One-Loop Effective Potentials With Many Scalars}'',
\newblock \href{http://dx.doi.org/10.1140/epjc/s10052-013-2588-2}{Eur. Phys. J.
  C\,\textbf{73},\,2588\,(2013)},
  \href{http://arxiv.org/abs/1307.1477}{arXiv:1307.1477}.

\bibitem{Athron:2014wta}
P.~Athron, J.-h. Park, D.~St\"ockinger, A.~Voigt,
\newblock ``{FlexibleSUSY \textendash{} a meta spectrum generator for
  supersymmetric models}'',
\newblock \href{http://dx.doi.org/10.1016/j.nuclphysbps.2015.09.413}{Nucl.
  Part. Phys. Proc.\,\textbf{273-275},\,2424\,(2016)},
  \href{http://arxiv.org/abs/1410.7385}{arXiv:1410.7385}.

\bibitem{Athron:2017fvs}
P.~Athron, M.~Bach, D.~Harries, T.~Kwasnitza, J.-h. Park, D.~St\"ockinger,
  A.~Voigt, J.~Ziebell,
\newblock ``{FlexibleSUSY 2.0: Extensions to investigate the phenomenology of
  SUSY and non-SUSY models}'',
\newblock \href{http://dx.doi.org/10.1016/j.cpc.2018.04.016}{Comput. Phys.
  Commun.\,\textbf{230},\,145\,(2018)},
  \href{http://arxiv.org/abs/1710.03760}{arXiv:1710.03760}.

\bibitem{Skands:2003cj}
P.~Z. Skands, \emph{et~al.},
\newblock ``{SUSY Les Houches accord: Interfacing SUSY spectrum calculators,
  decay packages, and event generators}'',
\newblock
  \href{http://dx.doi.org/10.1088/1126-6708/2004/07/036}{JHEP\,\textbf{07},\,036\,(2004)},
  \href{http://arxiv.org/abs/hep-ph/0311123}{arXiv:hep-ph/0311123}.

\bibitem{Allanach:2008qq}
B.~Allanach, \emph{et~al.},
\newblock ``{SUSY Les Houches Accord 2}'',
\newblock
  \href{http://dx.doi.org/10.1016/j.cpc.2008.08.004}{Comput.Phys.Commun.\,\textbf{180},\,8\,(2009)},
  \href{http://arxiv.org/abs/0801.0045}{arXiv:0801.0045}.

\bibitem{Feroz:2008xx}
F.~Feroz, M.~P. Hobson, M.~Bridges,
\newblock ``{MultiNest: an efficient and robust Bayesian inference tool for
  cosmology and particle physics}'',
\newblock \href{http://dx.doi.org/10.1111/j.1365-2966.2009.14548.x}{Mon. Not.
  Roy. Astron. Soc.\,\textbf{398},\,1601\,(2009)},
  \href{http://arxiv.org/abs/0809.3437}{arXiv:0809.3437}.

\bibitem{GAMBIT:2017yxo}
GAMBIT, P.~Athron, \emph{et~al.},
\newblock ``{GAMBIT: The Global and Modular Beyond-the-Standard-Model Inference
  Tool}'',
\newblock \href{http://dx.doi.org/10.1140/epjc/s10052-017-5321-8}{Eur. Phys. J.
  C\,\textbf{77},\,784\,(2017)},
  \href{http://arxiv.org/abs/1705.07908}{arXiv:1705.07908},
\newblock [Addendum: Eur.Phys.J.C 78, 98 (2018)].

\bibitem{GAMBITModelsWorkgroup:2017ilg}
GAMBIT Models Workgroup, P.~Athron, \emph{et~al.},
\newblock ``{SpecBit, DecayBit and PrecisionBit: GAMBIT modules for computing
  mass spectra, particle decay rates and precision observables}'',
\newblock \href{http://dx.doi.org/10.1140/epjc/s10052-017-5390-8}{Eur. Phys. J.
  C\,\textbf{78},\,22\,(2018)},
  \href{http://arxiv.org/abs/1705.07936}{arXiv:1705.07936}.

\bibitem{Bloor:2021gtp}
S.~Bloor, T.~E. Gonzalo, P.~Scott, C.~Chang, A.~Raklev, J.~E. Camargo-Molina,
  A.~Kvellestad, J.~J. Renk, P.~Athron, C.~Bal\'azs,
\newblock ``{The GAMBIT Universal Model Machine: from Lagrangians to
  likelihoods}'',
\newblock \href{http://dx.doi.org/10.1140/epjc/s10052-021-09828-9}{Eur. Phys.
  J. C\,\textbf{81},\,1103\,(2021)},
  \href{http://arxiv.org/abs/2107.00030}{arXiv:2107.00030}.

\bibitem{Martinez:2017lzg}
GAMBIT, G.~D. Martinez, J.~McKay, B.~Farmer, P.~Scott, E.~Roebber, A.~Putze,
  J.~Conrad,
\newblock ``{Comparison of statistical sampling methods with ScannerBit, the
  GAMBIT scanning module}'',
\newblock \href{http://dx.doi.org/10.1140/epjc/s10052-017-5274-y}{Eur. Phys. J.
  C\,\textbf{77},\,761\,(2017)},
  \href{http://arxiv.org/abs/1705.07959}{arXiv:1705.07959}.

\bibitem{Ren:2017ymm}
J.~Ren, L.~Wu, J.~M. Yang, J.~Zhao,
\newblock ``{Exploring supersymmetry with machine learning}'',
\newblock \href{http://dx.doi.org/10.1016/j.nuclphysb.2019.114613}{Nucl. Phys.
  B\,\textbf{943},\,114613\,(2019)},
  \href{http://arxiv.org/abs/1708.06615}{arXiv:1708.06615}.

\bibitem{Staub:2019xhl}
F.~Staub,
\newblock ``{xBIT: an easy to use scanning tool with machine learning
  abilities}'',
\newblock \,\,(2019), \href{http://arxiv.org/abs/1906.03277}{arXiv:1906.03277}.

\bibitem{SettlesReview}
{Settles, Burr},
\newblock ``{Active Learning Literature Survey}'',
\newblock \url{https://burrsettles.com/pub/settles.activelearning.pdf}.

\bibitem{Caron:2019xkx}
S.~Caron, T.~Heskes, S.~Otten, B.~Stienen,
\newblock ``{Constraining the Parameters of High-Dimensional Models with Active
  Learning}'',
\newblock \href{http://dx.doi.org/10.1140/epjc/s10052-019-7437-5}{Eur. Phys. J.
  C\,\textbf{79},\,944\,(2019)},
  \href{http://arxiv.org/abs/1905.08628}{arXiv:1905.08628}.

\bibitem{Rocamonde:2022gyw}
J.~Rocamonde, L.~Corpe, G.~Zilgalvis, M.~Avramidou, J.~Butterworth,
\newblock ``{Picking the low-hanging fruit: testing new physics at scale with
  active learning}'',
\newblock \,\,(2022), \href{http://arxiv.org/abs/2202.05882}{arXiv:2202.05882}.

\bibitem{Abdughani:2019wuv}
M.~Abdughani, J.~Ren, L.~Wu, J.~M. Yang, J.~Zhao,
\newblock ``{Supervised deep learning in high energy phenomenology: a mini
  review}'',
\newblock \href{http://dx.doi.org/10.1088/0253-6102/71/8/955}{Commun. Theor.
  Phys.\,\textbf{71},\,955\,(2019)},
  \href{http://arxiv.org/abs/1905.06047}{arXiv:1905.06047}.

\bibitem{Boehnlein:2021eym}
A.~Boehnlein, \emph{et~al.},
\newblock ``{Artificial Intelligence and Machine Learning in Nuclear
  Physics}'',
\newblock \,\,(2021), \href{http://arxiv.org/abs/2112.02309}{arXiv:2112.02309}.

\bibitem{Gili:2022oul}
K.~Gili, M.~Mauri, A.~Perdomo-Ortiz,
\newblock ``{Evaluating Generalization in Classical and Quantum Generative
  Models}'',
\newblock \,\,(2022), \href{http://arxiv.org/abs/2201.08770}{arXiv:2201.08770}.

\bibitem{DiversityAL}
Z.~Xu, R.~Akella, Y.~Zhang,
\newblock ``incorporating diversity and density in active learning for
  relevance feedback'',
\newblock in \emph{Advances in Information Retrieval}, edited by G.~Amati,
  C.~Carpineto, G.~Romano, pp. 246--257, Berlin, Heidelberg, 2007, Springer
  Berlin Heidelberg.

\bibitem{Goodfellow:2014upx}
I.~J. Goodfellow, J.~Pouget-Abadie, M.~Mirza, B.~Xu, D.~Warde-Farley, S.~Ozair,
  A.~Courville, Y.~Bengio,
\newblock ``{Generative Adversarial Networks}'',
\newblock \,\,(2014), \href{http://arxiv.org/abs/1406.2661}{arXiv:1406.2661}.

\bibitem{Mohamed:2016xyz}
S.~Mohamed, B.~Lakshminarayanan,
\newblock ``{Learning in Implicit Generative Models}'',
\newblock \,\,(2016), \href{http://arxiv.org/abs/1610.03483}{arXiv:1610.03483}.

\bibitem{Lopez:2022lkd}
M.~Lopez, V.~Boudart, K.~Buijsman, A.~Reza, S.~Caudill,
\newblock ``{Simulating Transient Noise Bursts in LIGO with Generative
  Adversarial Networks}'',
\newblock \,\,(2022), \href{http://arxiv.org/abs/2203.06494}{arXiv:2203.06494}.

\bibitem{Mescheder:2018xyz}
L.~Mescheder, A.~Geiger, S.~Nowozin,
\newblock ``{Which Training Methods for GANs do actually Converge?}'',
\newblock \,\,(2018), \href{http://arxiv.org/abs/1801.04406}{arXiv:1801.04406}.

\bibitem{Ellis:2002wv}
J.~R. Ellis, K.~A. Olive, Y.~Santoso,
\newblock ``{The MSSM parameter space with nonuniversal Higgs masses}'',
\newblock \href{http://dx.doi.org/10.1016/S0370-2693(02)02071-3}{Phys. Lett.
  B\,\textbf{539},\,107\,(2002)},
  \href{http://arxiv.org/abs/hep-ph/0204192}{arXiv:hep-ph/0204192}.

\bibitem{Ellis_2009}
J.~Ellis, K.~A. Olive, P.~Sandick,
\newblock ``update on the direct detection of dark matter in {MSSM} models with
  non-universal higgs masses'',
\newblock \href{http://dx.doi.org/10.1088/1367-2630/11/10/105015}{New Journal
  of Physics\,\textbf{11},\,105015\,(2009)}.

\bibitem{Ellis:2012aa}
J.~Ellis, K.~A. Olive,
\newblock ``{Revisiting the Higgs Mass and Dark Matter in the CMSSM}'',
\newblock \href{http://dx.doi.org/10.1140/epjc/s10052-012-2005-2}{Eur. Phys. J.
  C\,\textbf{72},\,2005\,(2012)},
  \href{http://arxiv.org/abs/1202.3262}{arXiv:1202.3262}.

\bibitem{Aghanim:2018eyx}
Planck, N.~Aghanim, \emph{et~al.},
\newblock ``{Planck 2018 results. VI. Cosmological parameters}'',
\newblock \href{http://dx.doi.org/10.1051/0004-6361/201833910}{Astron.
  Astrophys.\,\textbf{641},\,A6\,(2020)},
  \href{http://arxiv.org/abs/1807.06209}{arXiv:1807.06209}.

\bibitem{Griest:1989wd}
K.~Griest, M.~Kamionkowski,
\newblock ``{Unitarity Limits on the Mass and Radius of Dark Matter
  Particles}'',
\newblock \href{http://dx.doi.org/10.1103/PhysRevLett.64.615}{Phys. Rev.
  Lett.\,\textbf{64},\,615\,(1990)}.

\bibitem{Hedri:2014mua}
S.~El~Hedri, W.~Shepherd, D.~G.~E. Walker,
\newblock ``{Perturbative Unitarity Constraints on Gauge Portals}'',
\newblock \href{http://dx.doi.org/10.1016/j.dark.2017.09.006}{Phys. Dark
  Univ.\,\textbf{18},\,127\,(2017)},
  \href{http://arxiv.org/abs/1412.5660}{arXiv:1412.5660}.

\bibitem{vonHarling:2014kha}
B.~von Harling, K.~Petraki,
\newblock ``{Bound-state formation for thermal relic dark matter and
  unitarity}'',
\newblock
  \href{http://dx.doi.org/10.1088/1475-7516/2014/12/033}{JCAP\,\textbf{12},\,033\,(2014)},
  \href{http://arxiv.org/abs/1407.7874}{arXiv:1407.7874}.

\bibitem{Cahill-Rowley:2015aea}
M.~Cahill-Rowley, S.~El~Hedri, W.~Shepherd, D.~G.~E. Walker,
\newblock ``{Perturbative Unitarity Constraints on Charged/Colored Portals}'',
\newblock \href{http://dx.doi.org/10.1016/j.dark.2018.04.003}{Phys. Dark
  Univ.\,\textbf{22},\,48\,(2018)},
  \href{http://arxiv.org/abs/1501.03153}{arXiv:1501.03153}.

\bibitem{Kahlhoefer:2015bea}
F.~Kahlhoefer, K.~Schmidt-Hoberg, T.~Schwetz, S.~Vogl,
\newblock ``{Implications of unitarity and gauge invariance for simplified dark
  matter models}'',
\newblock
  \href{http://dx.doi.org/10.1007/JHEP02(2016)016}{JHEP\,\textbf{02},\,016\,(2016)},
  \href{http://arxiv.org/abs/1510.02110}{arXiv:1510.02110}.

\bibitem{Baldes:2017gzw}
I.~Baldes, K.~Petraki,
\newblock ``{Asymmetric thermal-relic dark matter: Sommerfeld-enhanced
  freeze-out, annihilation signals and unitarity bounds}'',
\newblock
  \href{http://dx.doi.org/10.1088/1475-7516/2017/09/028}{JCAP\,\textbf{09},\,028\,(2017)},
  \href{http://arxiv.org/abs/1703.00478}{arXiv:1703.00478}.

\bibitem{ElHedri:2017nny}
S.~El~Hedri, A.~Kaminska, M.~de~Vries, J.~Zurita,
\newblock ``{Simplified Phenomenology for Colored Dark Sectors}'',
\newblock
  \href{http://dx.doi.org/10.1007/JHEP04(2017)118}{JHEP\,\textbf{04},\,118\,(2017)},
  \href{http://arxiv.org/abs/1703.00452}{arXiv:1703.00452}.

\bibitem{ElHedri:2018atj}
S.~El~Hedri, M.~de~Vries,
\newblock ``{Cornering Colored Coannihilation}'',
\newblock
  \href{http://dx.doi.org/10.1007/JHEP10(2018)102}{JHEP\,\textbf{10},\,102\,(2018)},
  \href{http://arxiv.org/abs/1806.03325}{arXiv:1806.03325}.

\bibitem{Harz:2018csl}
J.~Harz, K.~Petraki,
\newblock ``{Radiative bound-state formation in unbroken perturbative
  non-Abelian theories and implications for dark matter}'',
\newblock
  \href{http://dx.doi.org/10.1007/JHEP07(2018)096}{JHEP\,\textbf{07},\,096\,(2018)},
  \href{http://arxiv.org/abs/1805.01200}{arXiv:1805.01200}.

\bibitem{Hektor:2019ote}
A.~Hektor, A.~Hryczuk, K.~Kannike,
\newblock ``{Improved bounds on $\mathbb{Z}_{3}$ singlet dark matter}'',
\newblock
  \href{http://dx.doi.org/10.1007/JHEP03(2019)204}{JHEP\,\textbf{03},\,204\,(2019)},
  \href{http://arxiv.org/abs/1901.08074}{arXiv:1901.08074}.

\bibitem{Kannike:2019mzk}
K.~Kannike, K.~Loos, M.~Raidal,
\newblock ``{Gravitational wave signals of pseudo-Goldstone dark matter in the
  $\mathbb{Z}_{3}$ complex singlet model}'',
\newblock \href{http://dx.doi.org/10.1103/PhysRevD.101.035001}{Phys. Rev.
  D\,\textbf{101},\,035001\,(2020)},
  \href{http://arxiv.org/abs/1907.13136}{arXiv:1907.13136}.

\bibitem{Alanne:2020jwx}
T.~Alanne, N.~Benincasa, M.~Heikinheimo, K.~Kannike, V.~Keus, N.~Koivunen,
  K.~Tuominen,
\newblock ``{Pseudo-Goldstone dark matter: gravitational waves and
  direct-detection blind spots}'',
\newblock
  \href{http://dx.doi.org/10.1007/JHEP10(2020)080}{JHEP\,\textbf{10},\,080\,(2020)},
  \href{http://arxiv.org/abs/2008.09605}{arXiv:2008.09605}.

\bibitem{Fuks:2020tam}
B.~Fuks, M.~D. Goodsell, D.~W. Kang, P.~Ko, S.~J. Lee, M.~Utsch,
\newblock ``{Heavy dark matter through the dilaton portal}'',
\newblock
  \href{http://dx.doi.org/10.1007/JHEP10(2020)044}{JHEP\,\textbf{10},\,044\,(2020)},
  \href{http://arxiv.org/abs/2007.08546}{arXiv:2007.08546}.

\bibitem{Espinoza:2020qyf}
C.~Espinoza, M.~Mondrag\'on,
\newblock ``{Prospects of Indirect Detection for the Heavy S3 Dark Doublet}'',
\newblock \,\,(2020), \href{http://arxiv.org/abs/2008.11792}{arXiv:2008.11792}.

\bibitem{Espinoza:2020kut}
C.~Espinoza, E.~Garc\'es, M.~Mondrag\'on, H.~Reyes-Gonz\'alez,
\newblock ``{An Inert Scalar In The S3 Symmetric Model}'',
\newblock \href{http://dx.doi.org/10.1088/1742-6596/1586/1/012025}{J. Phys.
  Conf. Ser.\,\textbf{1586},\,012025\,(2020)}.

\bibitem{Cao:2013wqa}
J.~Cao, P.~Wan, J.~M. Yang, J.~Zhu,
\newblock ``{The SM extension with color-octet scalars: diphoton enhancement
  and global fit of LHC Higgs data}'',
\newblock
  \href{http://dx.doi.org/10.1007/JHEP08(2013)009}{JHEP\,\textbf{08},\,009\,(2013)},
  \href{http://arxiv.org/abs/1303.2426}{arXiv:1303.2426}.

\bibitem{He:2013tla}
X.-G. He, H.~Phoon, Y.~Tang, G.~Valencia,
\newblock ``{Unitarity and vacuum stability constraints on the couplings of
  color octet scalars}'',
\newblock
  \href{http://dx.doi.org/10.1007/JHEP05(2013)026}{JHEP\,\textbf{05},\,026\,(2013)},
  \href{http://arxiv.org/abs/1303.4848}{arXiv:1303.4848}.

\bibitem{Cheng:2018mkc}
L.~Cheng, O.~Eberhardt, C.~W. Murphy,
\newblock ``{Novel theoretical constraints for color-octet scalar models}'',
\newblock \href{http://dx.doi.org/10.1088/1674-1137/43/9/093101}{Chin. Phys.
  C\,\textbf{43},\,093101\,(2019)},
  \href{http://arxiv.org/abs/1808.05824}{arXiv:1808.05824}.

\bibitem{Schuessler:2007av}
A.~Schuessler, D.~Zeppenfeld,
\newblock ``{Unitarity constraints on MSSM trilinear couplings}'',
\newblock in \emph{{SUSY 2007 Proceedings}}, pp. 236--239, 2007,
  \href{http://arxiv.org/abs/0710.5175}{arXiv:0710.5175}.

\bibitem{SchuesslerThesis}
{A. Sch\"ussler},
\newblock ``{Unitarit\"ats-Schranken an triskalare Kopplungen im MSSM}'',
\newblock {Diplomarbeit, Institut f\"ur Theoretische Physik, Universit\"at
  Karlsruhe}\,\,(2005),
  \href{https://www.itp.kit.edu/prep/diploma/PSFiles/Diplom_Schuessler.ps.gz}{KIT
  diploma server}.

\bibitem{Staub:2018vux}
F.~Staub,
\newblock ``{Theoretical Constraints on Supersymmetric Models: Perturbative
  Unitarity vs. Vacuum Stability}'',
\newblock \href{http://dx.doi.org/10.1016/j.physletb.2018.12.039}{Phys. Lett.
  B\,\textbf{789},\,203\,(2019)},
  \href{http://arxiv.org/abs/1811.08300}{arXiv:1811.08300}.

\bibitem{Baker:2020vkh}
M.~J. Baker, P.~Cox, R.~R. Volkas,
\newblock ``{Has the Origin of the Third-Family Fermion Masses been
  Determined?}'',
\newblock \,\,(2020), \href{http://arxiv.org/abs/2012.10458}{arXiv:2012.10458}.

\bibitem{Goodsell:2020lpx}
M.~D. Goodsell, S.~Kraml, H.~Reyes-Gonz\'alez, S.~L. Williamson,
\newblock ``{Constraining Electroweakinos in the Minimal Dirac Gaugino
  Model}'',
\newblock \href{http://dx.doi.org/10.21468/SciPostPhys.9.4.047}{SciPost
  Phys.\,\textbf{9},\,047\,(2020)},
  \href{http://arxiv.org/abs/2007.08498}{arXiv:2007.08498}.

\bibitem{Chalons:2018gez}
G.~Chalons, M.~D. Goodsell, S.~Kraml, H.~Reyes-González, S.~L. Williamson,
\newblock ``{LHC limits on gluinos and squarks in the minimal Dirac gaugino
  model}'',
\newblock
  \href{http://dx.doi.org/10.1007/JHEP04(2019)113}{JHEP\,\textbf{04},\,113\,(2019)},
  \href{http://arxiv.org/abs/1812.09293}{arXiv:1812.09293}.

\bibitem{Miller:2020hua}
B.~K. Miller, A.~Cole, G.~Louppe, C.~Weniger,
\newblock ``{Simulation-efficient marginal posterior estimation with swyft:
  stop wasting your precious time}'',
\newblock \,\,(2020), \href{http://arxiv.org/abs/2011.13951}{arXiv:2011.13951}.

\end{thebibliography}

\end{document}